\documentclass[a4paper,11pt]{article}
\usepackage{AuxiliaryFiles/aaskaiid}
\usepackage{AuxiliaryFiles/orcidlink}
\usepackage[rightcaption]{sidecap}
\usepackage{subcaption}
\title{The Cosmic Radio Dipole and Observer Effects in SKA all-sky surveys}
\ShortTitle{The Cosmic Radio Dipole}

\ShortName{SKA SWG Cosmology}
\author[1,2,3]{D.~Bertacca\orcidlink{0000-0002-2490-7139}}
\author[4]{L.~Böhme\orcidlink{0009-0007-2702-2479}}
\author[1,2]{B.~Bottazzi~Baldi\orcidlink{0009-0002-9754-9456}}
\author[5,6]{C.~Burigana\orcidlink{0000-0002-3005-5796}}
\author[7,8,9,10]{M.~Y.~Elkhashab\orcidlink{0000-0001-9306-2603}}
\author[11,12]{C.~L.~Hale\orcidlink{0000-0001-6279-4772}}
\author[13,14]{S.~von~Hausegger\orcidlink{0000-0002-6274-1424}}
\author[15]{H.-R.~Kl\"ockner\orcidlink{0000-0002-0648-2704}}
\author[16]{M.~Land-Strykowski\orcidlink{0009-0002-1785-1699}}
\author[16]{G.~F.~Lewis\orcidlink{0000-0003-3081-9319}}
\author[16]{V.~Mittal\orcidlink{0000-0002-1708-6088}}
\author[16]{O.~T.~Oayda\orcidlink{0009-0002-5013-8959}}
\author[4]{D.~J.~Schwarz\orcidlink{0000-0003-2413-0881}}
\author[17]{P.~Tiwari\orcidlink{0000-0001-7888-4270}}
\author[5,6]{T.~Trombetti\orcidlink{0000-0001-5166-2467}}
\author[15]{J.~D.~Wagenveld\orcidlink{0000-0003-1321-0886}}

\affiliation[1]{Dipartimento di Fisica e Astronomia Galileo Galilei, Università di Padova, 35131, Padova, Italy}
\affiliation[2]{INFN Sezione di Padova, I-35131, Padova, Italy}
\affiliation[3]{INAF, Osservatorio Astronomico di Padova, Italy}
\affiliation[4]{Fakultät für Physik, Universität Bielefeld, Postfach 100131, 33501 Bielefeld, Germany}
\affiliation[5]{INAF, Istituto di Radioastronomia, Via Piero Gobetti 101, 40129 Bologna, Italy}
\affiliation[6]{INFN, Sezione di Bologna, Via Irnerio 46, 40127 Bologna, Italy}
\affiliation[7]{Dipartimento di Fisica – Sezione di Astronomia, Università di Trieste, Via Tiepolo 11, 34131, Trieste, Italy}
\affiliation[8]{INAF, Osservatorio Astronomico di Trieste, Via Tiepolo 11, I-34131 Trieste, Italy}
\affiliation[9]{INFN, Sezione di Trieste, Via Valerio 2, 34127 Trieste TS, Italy}
\affiliation[10]{IFPU, Institute for Fundamental Physics of the Universe, via Beirut 2, 34151 Trieste, Italy}
\affiliation[11]{Institute for Astronomy, Royal Observatory Edinburgh, Blackford Hill, Edinburgh, EH9 3HJ, UK}
\affiliation[12]{Astrophysics, Department of Physics, University of Oxford, Keble Road, Oxford, OX1 3RH, UK}
\affiliation[13]{Department of Physics, Yale University, 217 Prospect St, New Haven, CT 06511, USA}
\affiliation[14]{Rudolf Peierls Centre for Theoretical Physics, University of Oxford, Parks Road, Oxford OX1 3PU, UK}
\affiliation[15]{Max-Planck-Institut f\"ur Radioastronomie, Auf dem H\"ugel 69, 53121 Bonn, Germany}
\affiliation[16]{Sydney Institute for Astronomy, School of Physics A28, The University of Sydney, NSW 2006, Australia}
\affiliation[17]{Department of Physics, Guangdong Technion - Israel Institute of Technology, Shantou, Guangdong 515063, P.R. China}

\emailAdd{sebastian.vonhausegger@physics.ox.ac.uk}
\emailAdd{dschwarz@physik.uni-bielefeld.de}

\def\lsim{\,\lower2truept\hbox{${<\atop\hbox{\raise4truept\hbox{$\sim$}}}$}\,}
\def\gsim{\,\lower2truept\hbox{${> \atop\hbox{\raise4truept\hbox{$\sim$}}}$}\,}

\abstract{The Square Kilometre Array (SKA) will observe our Universe on the largest scales and so enable the study of fundamental assumptions in cosmology, such as cosmic homogeneity and isotropy.  Observations over recent years have reaffirmed an unexpected challenge to the Cosmological Principle first seen in measurements of the cosmic radio dipole. This dipole anisotropy is thought to be caused by our local observer motion, however, it is observed to be about three times larger than predicted, with significances exceeding 5$\sigma$. The SKA is in a unique position to shed light on this anomaly by measuring this and other effects related to our observer motion. To quantify this, we forecast the capability of large-angle HI and continuum surveys of the SKA to constrain the kinematic matter dipole in galaxy number counts, the modulation of the matter power spectrum by a kinematic dipole, and kinematic spectral distortions of diffuse radio emission.}

\begin{document}

\maketitle
\tableofcontents

\newpage
\section{Introduction}
\label{sec:introduction}

The SKA-Low and the SKA-Mid instruments will map the radio sky over wide areas with exquisite survey speed and depth. In particular, they will perform radio continuum measurements of extragalactic sources, which, in contrast to, e.g., optical surveys, are unaffected by Galactic dust extinction. Thus the SKA surveys are ideal to address questions about the largest scales of the Universe, accessing scales that span 180 degrees on the sky, and probe linear comoving distances of up to $\sim10\,{\rm Gpc}$ and redshifts well beyond unity.
Alongside other modern radio continuum surveys, the SKA will detect active galactic nuclei out to the redshifts of their birth and probe the population of star-forming galaxies, which offers interesting possibilities of single- and multi-tracer studies of the cosmic large-scale structure.

In this contribution we will focus on the largest accessible scales that appear as a dipole on the sky. As such, the study of cosmic dipoles relates directly to foundational tests of modern cosmology, which is based on assuming large-scale homogeneity and isotropy.  This target poses challenging requirements on instrument calibration and systematics control, such as flux density calibration and homogeneous survey depth.  Ensuring these will also benefit science cases other than the one presented here~\citep[e.g.][]{Asorey01.2026.SKA,Harrison01.2026.SKA}. Simultaneously, the proposed cosmological tests require a solid understanding of the astrophysics of radio sources of all kinds.  In turn, the present science case will benefit from the results obtained by other working groups' studies --- extragalactic and galactic studies alike~\citep[e.g.][]{Ronconi01.2026.SKA} --- especially those on local, diffuse foregrounds which might affect estimates of power and signals on large scales~\citep[e.g.][]{Spinelli01.2026.SKA}.  Finally, the exploration of physics on the largest scales is shared also with efforts to constrain large-scale, primordial non-Gaussianity, or effects beyond general relativity~\citep{Camera01.2026.SKA}.  A selection of ideas on how the SKAO instruments will lead to potentially transformational insights has already been presented in the previous Science Book~\citep{bourke2015advancing} and the SKA Cosmology Red Book~(\citet{SKA:2018ckk}, wherein \citet{Schwarz:2018swk}/\citet{Bengaly:2018ykb} focus on the radio dipole specifically). Here we report on further developments in the field.

The particular studies in this contribution, while arguably valuable in their own regard, are motivated by what is now referred to as ``the cosmic dipole anomaly''.  The cosmic dipole anomaly is the observation of a significant excess dipole in the number counts of cosmologically distant sources, beyond the signal expected due to our (heliocentric) observer motion with respect to the cosmic rest frame (CRF). The CRF is conventionally assumed to be that in which the prominent dipole of the CMB vanishes (also called the CMB rest frame), and in the standard cosmological model is also that of the matter field. The dipole anomaly therefore directly challenges this assumption.  While only hints of this anomaly (then called the radio dipole anomaly) existed back when the last SKA forecasts were undertaken~\citep[e.g.~][]{Blake:2002gx,Singal:2011dy,Rubart:2013tx,Tiwari:2014ni,Tiwari:2016adi},
it has since been established at high significance with additional data -- at $4.9\sigma$ for the first time with number counts of infrared sources~(\cite{Secrest:2020has,Secrest:2022uvx}, see also~\cite{Dam:2022wwh}) using data from WISE, and at $4.8\sigma$ and $5.4\sigma$ with number counts again of radio sources, using different combinations of data sets from the VLA, ASKAP, and LOFAR~\citep[][respectively]{Wagenveld:2023kvi,Bohme:2025nvu} --- see~\cite{Secrest:2025wyu} for a review of these developments.
All these observations have in common that they probe the large-scale spatial distribution of active galaxies, mostly radio galaxies and quasars, and strongly suggest that the largest cosmological scales are a very valuable target for SKA surveys.

An explanation for this anomaly is still outstanding.  A central question is whether the excess dipole is fully kinematic (i.e.~an apparent imprint on a largely isotropic background, caused by our observer motion) or whether it is due to unexpected clustering on large scales (i.e.~unexpected deviations from isotropy).  Intriguingly, the observed dipoles' directions seem to roughly agree with that of the CMB dipole (within a margin of the order of $10^\circ$), which suggests a common, possibly kinematic, origin. At face value, a kinematic origin would suggest that the sources collectively move with respect to the CMB rest frame, in the direction opposite to the Sun's motion. This curious finding ought to be seen in the light of the still unexplained anomalies of the CMB anisotropies at the largest angular scales~\citep{Schwarz:2015cma}, which exhibit alignments among the orientations of dipole, quadrupole and octupole. Some initial attempts have been made to distinguish kinematic from clustering contributions to the anomalous source count dipoles~(\cite{Wagenveld:2025ewl}; for another proposed approach, see~\citet{Nadolny:2021hti}), or to detect imprints of our observer motion with respect to the matter field also on other observables~\citep{daSilveiraFerreira:2024ddn,Tiwari:2024SDSS}, although a clear verdict is still outstanding. In addition the influence of different source populations and instrumental systematics has been studied in terms of gravitational clustering alone \citep[e.g.][]{Tiwari:2022hnf,Oayda:2024hnu,vonHausegger:2025iuo} \citep[see, also, ][]{Rubart:2014lia} as and of tensions between the different measurements~\citep{Land-Strykowski:2025gkz,Land-Strykowski:2026hmv}.  By measuring the large-scale modes of radio source counts as well as kinematic effects on a number of observables, the SKA therefore offers the opportunity to conclusively address current puzzles surrounding these cosmic anomalies.

In the present study we will focus on three observables and demonstrate the capabilities of SKA-Low and SKA-Mid continuum and HI surveys, in their AA4 configuration, to measure them.  Throughout, AA4 refers to the full design-baseline array assembly, while AA$^*$ denotes the reduced initial deployment.  Sec.~\ref{sec:numbercountsdipole} summarizes the design and results of forecasting inferences of large-scale multipoles in the number density of radio sources, with particular focus on the kinematic matter dipole.  Sec.~\ref{sec:psdipole} describes how to constrain our observer motion using the correspondingly dipole-modulated three-dimensional power spectrum.  Finally, Sec.~\ref{sec:spectrumdipole} considers (kinematic) spectral distortions of radio backgrounds that too exhibit anisotropies indicative of our motion with respect to these backgrounds.  Each section bases its forecasts on distinct simulations for the different investigated phenomena; their common basis (the charted sensitivity of the SKA instruments and assumptions regarding observing time etc.), however, is shared by all sections.  We summarize in Sec.~\ref{sec:summary}.

\section{Kinematic effects on number counts -- The Cosmic Radio Dipole}
\label{sec:numbercountsdipole}

A boosted observer sees the CMB temperature modulated in a number of ways, first and foremost by the appearance of a prominent Doppler dipole of amplitude $\mathcal{D}_{\rm CMB} = \beta$, relative to the CMB monopole.  Kinematic effects also modulate observables of other cosmological tracers such as number counts of galaxies, which is the subject of this section.  Using bespoke simulations of continuum radio galaxies we forecast the capabilities of the SKA to detect the apparent kinematic number count dipole along with measurements of other large-scale modes.

Since the initial hints~\citep[e.g.][]{Blake:2002gx,Singal:2011dy,Rubart:2013tx} of a matter dipole anomaly in the radio, measurements of the matter dipole also at other wavelengths have yielded significant results~\citep[e.g.][]{Secrest:2020has,Dam:2022wwh,vonHausegger:2025iuo}.  Nevertheless, the radio sky presents a potentially cleaner view of the background sources needed for a dipole measurement. This is not only due to the fact that there is no dust obscuration (Galactic or otherwise), but there are also very few compact sources that appear as part of the Galactic foreground outside the Galactic plane (stars being a primary concern) that emit appreciably in the radio, which means that only minimal selections have to be made for a clean sample. Furthermore, the population of bright ($\gtrsim10$ mJy at $\sim1\,{\rm GHz}$) radio sources is dominated by active galactic nuclei (AGN) at cosmological distances ($z\sim1$), presenting an ideal background sample. Most measurements of the radio dipole thus far have probed this population, which yields at most a few $10^5$ sources at the flux density cuts typically applied. The SKA surveys will readily surpass the depth of currently available surveys (for an overview, see Sect.~3 of~\cite{Asorey01.2026.SKA}), allowing for more accurate dipole measurements through sheer numbers. However, a sufficiently deep survey also probes the fainter population of star-forming galaxies (SFGs), the bright end of which presents a large population of low-redshift interlopers which can impart the signal of local clustering onto the measured dipole. In the context of radio dipole measurements, a significant number of SFGs has so far only featured in~\citet{Wagenveld:2024qhn}, calling for dedicated studies with upcoming data. As we demonstrate below, additional redshift information is required to remove the influence of local SFG clustering so as to ensure a reliable measurement of the cosmic radio dipole.

\subsection{Theory}
\label{sec:numbercountsdipole_theory}

The kinematic number count dipole appears in the projected number counts of a flux limited catalogue primarily due to two additive effects: \textit{i)}~special relativistic aberration of the angular area element modulates the number density of sources with a dipole aligned with the boost direction of relative amplitude $2\,\beta$, and \textit{ii)}~individual sources' flux densities are boosted past the applied flux limit along the same axis and hence further dipole-modulate the sample's number density.  \citet{Ellis:1984uka} determined that, overall, the number density of sources with spectra characterized by spectral indices $\alpha_i$ and with spectral flux densities above a given flux cut, $S_i(\nu)>S_*$, will appear modulated by a dipole of amplitude
\begin{align}
    \mathcal{D}_{\rm kin}(S_*) = \left[2+x(S_*)(1+\alpha(S_*))\right]\,\beta,
    \label{eq:Dkin_EB}
\end{align}
in the direction of the observer velocity, $\boldsymbol{\varv}_{\rm obs}$, towards the rest frame of these sources, and where $\beta\equiv |\boldsymbol{\varv}_{\rm obs}|/c$.  $\mathcal{D}_{\rm kin}$ is the so-called kinematic matter dipole amplitude whose measurement with mock data is described in the following subsections.

In Eq.~\eqref{eq:Dkin_EB}, $x$ denotes the slope of the full sample's integral source counts,
\begin{align}
    x(S_*)\equiv -\frac{{\rm d}\log}{{\rm d}\log S_*}\frac{{\rm d}N}{{\rm d}\Omega}(S>S_*),
\end{align}
which equates to the magnification bias of the considered sample, averaged over its redshift distribution, and $\alpha_i$ denotes source $i$'s spectral index in the observed frequency band\footnote{We read the sources' $\alpha$ from the T-RECS simulation, by interpolating the fluxes at the frequencies neighbouring $780\,{\rm MHz}$, i.e.~we do not include here variations introduced by the band's frequency response, nor do we recompute the spectral index from the observed (noisy) flux densities.}
\begin{align}
    \alpha_i(\nu)=-\frac{{\rm d}\log}{{\rm d}\log\nu}S_i(\nu).
\end{align}
The accurate computation of the expected dipole amplitude given $\beta$, Eq.~\eqref{eq:Dkin_EB}, contains a subtlety in that both $\alpha$ and $x$ are to be evaluated at the flux density cut $S_*$: $\alpha$ is the average spectral index of those sources with flux density close to $S_*$ and $x$ is measured only at the chosen $S_*$ of the considered catalogue.  This renders the kinematic dipole expectation a function of the applied flux density cut, $S_*$, as well as of the sample properties~\citep{vonHausegger:2024jan}.

While Eq.~\eqref{eq:Dkin_EB} presumes projected number counts integrated over all redshifts, predictions for the kinematic matter dipole amplitude as a function of redshift~\citep{Maartens:2017qoa} are mapped to observations by computing the above quantities in redshift bins~\citep{vonHausegger:2024fcu}.  Here, however, terms in addition to those in the above expression arise due to the Doppler boosting of source redshifts themselves, resulting in a dependency also on the slope of the redshift distribution at the bin's boundaries, e.g.
\begin{align}
    \mathcal{D}_{\rm kin}(z_b,S_*) = \left[2+x(z_b,S_*)(1+\alpha(z_b,S_*))+\left.(1+z)f_b(z)\right|^{z_2}_{z_1}\right]\,\beta,
    \label{eq:Dkin_z}
\end{align}
where $f_b(z)$ denotes the redshift distribution normalized to the redshift bin, $z_b$, here considered to be a top-hat selection with $z\in[z_1,z_2]$ (for a general expression, also allowing for photometric redshift measurements see equations~20 and 21 in~\citet{vonHausegger:2024fcu}).  In particular, dipole measurements in samples where low-redshift sources, e.g.~$z\lesssim0.1$, have been removed via redshift cuts require such corrections for an accurate~\citeauthor{Ellis:1984uka} test.

\subsection{Mocks}
\label{sec:numbercountsdipole_mocks}

We generate mock observations of sources (SFG and AGN) at the nominal SKA-Mid Band 1 observing frequency of $\nu=780\,{\rm MHz}$, sampled from a best-fit $\Lambda$CDM matter density field~\citep[using \texttt{glass}, \texttt{camb}, and \texttt{CCL}; ][]{Tessore:2023zyk,Lewis:1999bs,LSSTDarkEnergyScience:2018yem}, that contain expected source populations, properties and number densities generated from the T-RECS simulation~\citep{Bonaldi:2023fnb}, and account for survey-specific noise and source detectability following the SKAO sensitivity calculator for AA4 and SKA-Mid, assuming $10\,000$ hours of observation distributed across individual pointings covering ${\rm Dec}<0$~\citep[following the specifications in ][]{SKA:2018ckk}.  The base simulations are the same as those used for the continuum forecasts reported in \citet{Asorey01.2026.SKA}, and we refer to their section 5 for full details on the mock generation.  

For the present purpose, however, we modify these mocks by injecting the expected kinematic (dipole) signal, pointed at $({\rm RA},{\rm Dec})=(167.942^\circ, -6.944^\circ)$ with $\beta=1.23\times10^{-3}$~\citep{Planck:2018nkj}, by aberrating the source positions and Doppler modulating the sources' intrinsic spectral flux densities, $S'_i\approx(1+\beta\cos\theta_i)^{1+\alpha}S_i$ and intrinsic (cosmological) redshifts, $z'_i\approx(1+z_i)(1-\beta\cos\theta_i)-1$, accordingly.  We aberrate the source positions prior to computing the detector sensitivity at the sampled position, and also introduce the corresponding Doppler modulation prior to introducing noise.  Finally, to simulate photometric redshift measurements we further sample observed redshifts from Gaussian uncertainties around the boosted redshifts as $z_{i,\rm obs}\sim\mathcal{G}(z'_i,\sigma_z\,(1+z'_i))$ with $\sigma_z=0.03$ for AA4~\citep{Harrison:2016stv}.

As in \citet{Asorey01.2026.SKA}, we consider final catalogues after applying a range of cuts on $S_{\nu,{\rm obs}}$, the observed spectral flux density, as well as on source-by-source signal-to-noise ratio (SNR).  To highlight the impact of said low-redshift clustering we consider two sets of samples, once with and once without a redshift cut that excludes $z_{\rm obs}<0.5$.  Finally, we apply masks to exclude from our final maps the Galactic plane defined by Galactic latitude,~$|b|<10^\circ$.  It is these final catalogues for which we can compute the values of the average magnification bias $x$ and the average spectral index $\alpha$ as described above.  Tables~\ref{tab:mocks1} and~\ref{tab:mocks2} list the total number of sources in each of the produced samples along with the measured values of $x$ and $\alpha$ which determine the kinematic matter dipole expectation $\mathcal{D}_{\rm kin}$.\footnote{While the dipole expectations for the samples in Table~\ref{tab:mocks1} are set by Eq.~\eqref{eq:Dkin_EB}, the cut on photometric redshift $z_{\rm obs}$ defining the samples in Table~\ref{tab:mocks2} requires the generalization of Eq.~\eqref{eq:Dkin_z}, i.e.~equations 20/21 in \citet{vonHausegger:2024fcu}.} The rms dipole amplitude, $\sigma_{\mathcal{D},{\rm cl}}$, expected from $\Lambda$CDM clustering for this sample's redshift distribution and bias is computed directly from the mock's input theory power spectrum. Already here we find that the samples' kinematic dipole measurements are not shot-noise limited but instead are dominated by the rms expected from clustering, alleviated by brighter flux density cuts, or, more effectively, by higher redshift cuts.

\begin{table}[htbp]
    \centering
    \begin{tabular}{c|r|c|c|c|c}
        $S_*\;[\mu{\rm Jy}]$    &   \multicolumn{1}{c|}{$N$} &   $x$ &   $\alpha$    &   $\mathcal{D}_{\rm kin}$   &   $\sigma_{\mathcal{D},{\rm cl}}$\\\hline\hline
        40	&	88\,780\,182	&	1.18	&	0.79	&	5.06$\;\times10^{-3}$	&	2.00$\;\times10^{-3}$ 	\\
        60	&	54\,747\,714	&	1.21	&	0.79	&	5.12$\;\times10^{-3}$	&	2.39$\;\times10^{-3}$ 	\\
        80	&	38\,574\,098	&	1.23	&	0.79	&	5.16$\;\times10^{-3}$	&	2.84$\;\times10^{-3}$ 	\\
    \end{tabular}
    \caption{Total number of sources $N$ in our simulated sample (for details see~\citet{Asorey01.2026.SKA}) along with measured values for $x$, $\alpha$ and the kinematic matter dipole expectation $\mathcal{D}_{\rm kin}$ (see Sec.~\ref{sec:numbercountsdipole_theory}) for different choices of flux limits $S_*$ and after masking. A signal-to-noise ratio cut of 5 has been applied to all samples.  
    Samples contain sources with redshifts $z_{\rm obs}>0$.}
    \label{tab:mocks1}
\end{table}

\begin{table}[htbp]
    \centering
    \begin{tabular}{c|r|c|c|c|c}
        $S_*\;[\mu{\rm Jy}]$    &   \multicolumn{1}{c|}{$N$} &   $x$ &   $\alpha$    &   $\mathcal{D}_{\rm kin}$   &   $\sigma_{\mathcal{D},{\rm cl}}$\\\hline\hline
        40	&	59\,749\,189	&	1.32	&	0.81	&	3.24$\;\times10^{-3}$	&	0.92$\;\times10^{-3}$ 	\\
        60	&	34\,732\,062	&	1.36	&	0.81	&	3.21$\;\times10^{-3}$	&	0.87$\;\times10^{-3}$ 	\\
        80	&	23\,463\,951	&	1.37	&	0.81	&	3.23$\;\times10^{-3}$	&	0.94$\;\times10^{-3}$ 	\\
    \end{tabular}
    \caption{Same as Table~\ref{tab:mocks1} but for $z_{\rm obs}>0.5$. Here the kinematic matter dipole expectations have been corrected for the redshift boundary terms of \citet{vonHausegger:2024fcu}.}
    \label{tab:mocks2}
\end{table}

\subsection{Methods}
\label{sec:numbercountsdipole_methods}

After casting the sources into \texttt{HEALPix} pixels~\citep{Gorski:2004by}, $p$, and defining an appropriate mask, covering Galactic plane and regions around input sources (see, e.g., right panels of Fig.~3 in~\citet{Asorey01.2026.SKA}), we may model the observed number of sources in each pixel, $N_{\rm obs}=N_{\rm obs}(p)$, as a monopole $\overline{N}$ modulated by a select range of higher multipoles,
\begin{align}
    N_{\rm obs} \sim \lambda(\boldsymbol\varpi)= \overline{N}\times\left(1+\mathcal{D}\,\cos\theta\,+\,
    \textcolor{gray}{
    \begin{pmatrix}
    \;\textrm{higher-order}\;\\
    \textrm{multipoles}
    \end{pmatrix}
    }
    \right)\times\left(\frac{\sigma}{\sigma_0}\right)^{-p}
    ,\label{eq:numbercounts}
\end{align}
where $\lambda$ is the expected number of sources according to the model and $\boldsymbol\varpi$ lists any free parameters given the model $\lambda$.  $\mathcal{D}$ denotes the dipole amplitude (defined relative to the monopole as in Eq.~\eqref{eq:Dkin_EB}) and $\theta=\theta(p)$ denotes the angular distance between the centre of pixel $p$ and the dipole direction.  Expressions for the higher-multipole templates can be found in, e.g.,~\citet{Oayda:2024voo,vonHausegger:2025iuo}, which, when employed, can fully account for the cross-template correlation against which \citep[e.g.][]{Abghari:2024eja} warned.
For present data, ``clustering'' multipole contributions simply enter at the noise level, so that these as well as their correlations are considered comparably small, or are effectively absorbed into the explicitly modelled multipoles.  For their limited sky fraction as well as for their low-redshift populations, it will be important to model the SKA's source catalogues including such higher clustering multipoles, although we here refrain from an extended such study.  Instead, we consider the influence of the rms of a given pixel on its recorded number of sources, as this too leads to an anisotropic modulation of the number count signal\footnote{For a map of the expected rms per pointing see, e.g., Fig.~2 of \citet{Asorey01.2026.SKA}.}.  Following \citet{Wagenveld:2023kvi}, we parametrise this as a sensitivity dependence of the number counts to the per-pixel rms, $\sigma$, compared against a reference rms, $\sigma_0$, here chosen to be the median of the unmasked rms pixels, and with a free power law index $p$, cf.~last factor in Eq.~\eqref{eq:numbercounts}.

An unbiased inference of a signal like this relies on the definition of a likelihood appropriate for the data.  Recent studies~\citep[e.g.~][]{Siewert:2019poc,Pashapour-Ahmadabadi:2025ghj} have shown that projected number counts of radio sources are more dispersed than if they had been drawn from a Poisson distribution.  Consequently, the likelihoods employed for inferring low-multipole signals in our maps ought to reflect their underlying statistics as well.  The Poisson likelihood considered in this context previously~\citep[e.g.][]{Dam:2022wwh,Wagenveld:2023kvi,Mittal:2023xub} has therefore been replaced by a negative binomial likelihood~\citep{Bohme:2025nvu} and a generalized Poisson likelihood~\citep{vonHausegger:2025iuo}, both inspired by earlier work on galaxy clustering~\citep[e.g.~][]{1984ApJ...276...13S,1995MNRAS.274..213S,Kitaura:2013cwa}.  Both return equivalent results~\citep{vonHausegger:2025iuo}, and we therefore employ the generalized Poisson likelihood in the following, for its relative computational efficiency:
\begin{align}
    \ln \mathcal{L}(N_{\rm obs}|\lambda(\boldsymbol\varpi),b) =& \,\ln\left(\lambda(1-b)\right) + (N_{\rm obs}-1)\ln\left(\lambda(1-b)+N_{\rm obs}b\right) \nonumber\\
    &- \ln(N_{\rm obs}!) - \lambda(1-b) - N_{\rm obs}b,\label{eq:lnl_genpoisson}
\end{align}
where the additional parameter $b$ scales the degree of the overdispersion ($b=0$ returns a Poisson distribution).  We perform the inference with nested sampling as described in~\cite{Dam:2022wwh} and compute Bayesian evidences for quantitative model comparison. 
We use a Nested Sampling Monte Carlo algorithm \textsc{MLFriends} \citep{2016S&C....26..383B, 2019PASP..131j8005B}, which efficiently samples likelihood shells for complex multi-dimensional likelihood space.
This algorithm has been implemented in the \textsc{UltraNest}\footnote{\url{https://johannesbuchner.github.io/UltraNest/}} \textsc{python} package \citep{2021JOSS....6.3001B}.

\subsection{Forecasts}
\label{sec:numbercountsdipole_forecasts}

\begin{figure}
    \centering
    \begin{minipage}{0.49\textwidth}
        \centering
        \includegraphics[width=\linewidth]{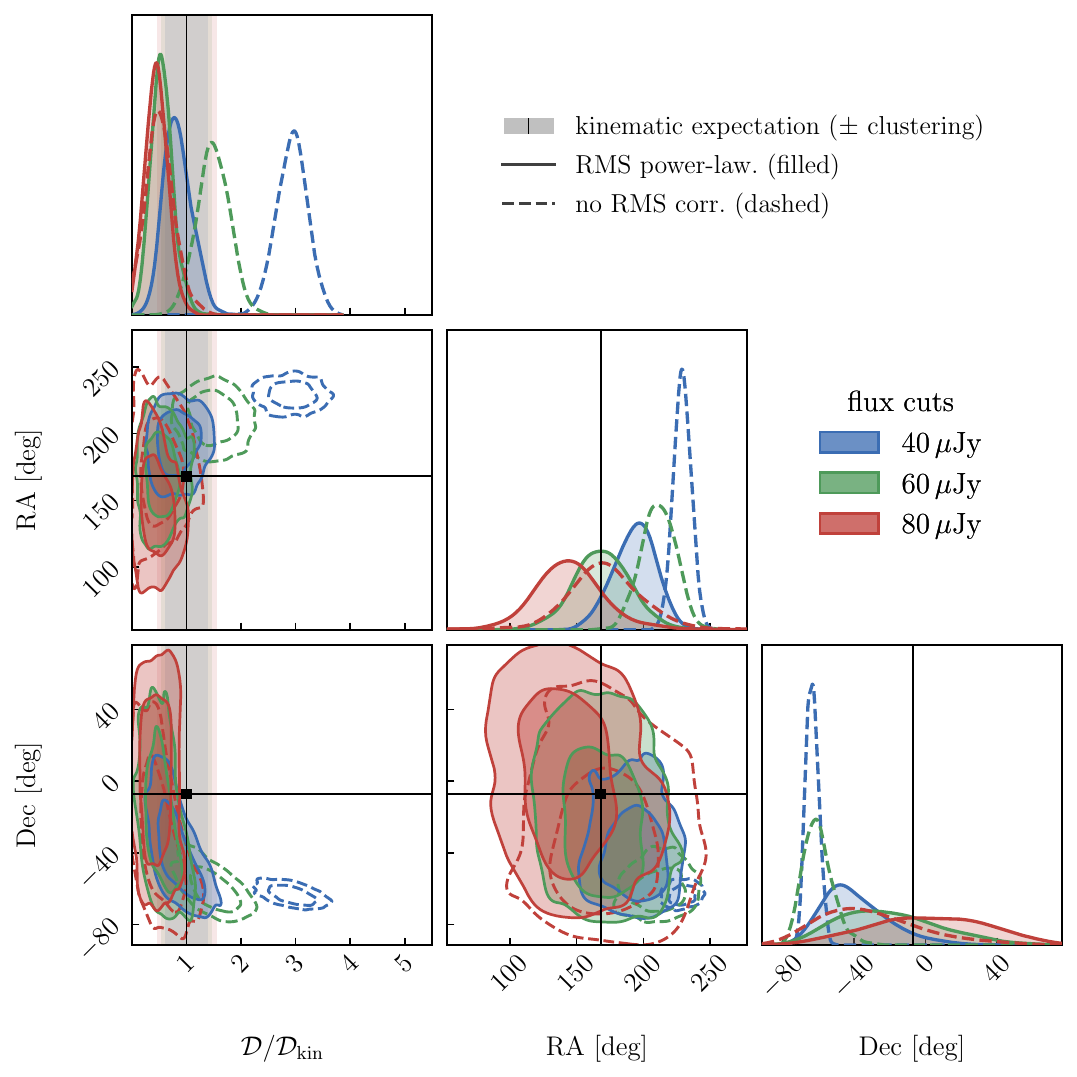}\\
        {\small \hspace{1.4em}$z > 0.0$}
    \end{minipage}\hfill
    \begin{minipage}{0.49\textwidth}
        \centering
        \includegraphics[width=\linewidth]{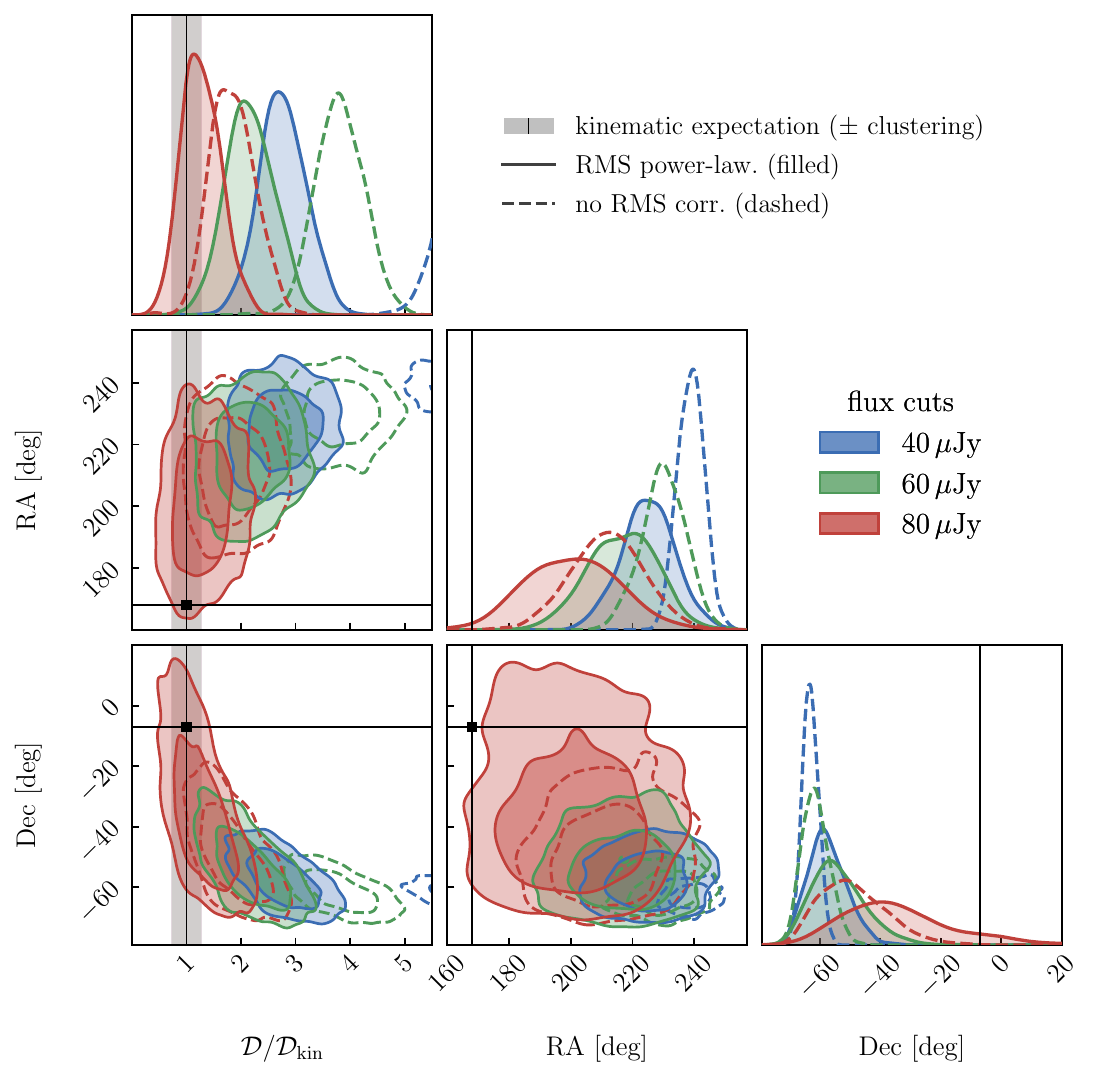}\\
        {\small \hspace{1.6em}$z_{\rm obs} > 0.5$}
    \end{minipage}
    \caption{Posteriors of the inferred free dipole parameters in our SKA mock using cuts corresponding to the samples in tables~\ref{tab:mocks1} and~\ref{tab:mocks2}.  All other model parameters are marginalized over for clarity.  The left panel uses all sources and the right panel only selects sources with $z_{\rm obs}>0.5$. Solid, filled contours show posteriors including the RMS correction, Eq.~\eqref{eq:numbercounts}, while dashed, empty contours show those without, where $p=0$.  We normalize the dipole amplitude by its kinematic expectation. We also show for reference spread of the expected clustering rms as shaded bands around the amplitude's expectation.} 
    \label{fig:posteriors1}
\end{figure}

We project the source positions onto a \texttt{HEALPix} grid with $N_{\rm side}=64$ and mask pixels $p$ appropriately to obtain a map $N_{\rm obs}(p)$.  Model parameters $\boldsymbol\varpi$ are then inferred from this map using the likelihood in Eq.~\eqref{eq:lnl_genpoisson}. We show the posteriors corresponding to a model with free monopole and dipole in Fig.~\ref{fig:posteriors1} where we present the results for our mock with and without our selected redshift cut $z_{\rm obs}>0.5$, with and without RMS correction, and for different flux density cuts $S>40,\,60,{\rm and}\,80\,\mu{\rm Jy}$.  As described above, a given sample is expected to contain relatively more sources at lower redshifts the fainter the flux cut. In those samples, as well as those without explicit redshift selection, we therefore expect to find a stronger contribution of the clustering dipole, making unambiguous detection of the sought-after cosmic matter dipole more difficult.
Indeed, our results demonstrate that the kinematic dipole is recovered best in samples with fainter and lower-redshift sources removed, and RMS corrections included in the model.

In both panels, and especially for the fainter flux density cuts, the RMS correction influences the inferred dipole parameters, brings the inferred dipole amplitudes and directions closer to their kinematic expectations in all cases, although even then the directions remain $16^\circ$ to $67^\circ$ away. The observed deviations are owed to the anisotropy in the RMS map itself, which effectively modulates the depth and source detection efficiency of the survey and thereby the number counts in a direction-dependent manner. An RMS correction on a purely {\it noise-limited} sample expects $p\approx x$, so values around $1.2$ to $1.4$ (see Tables~\ref{tab:mocks1} and~\ref{tab:mocks2}, and \citet{Wagenveld:2023kvi}), and on a purely {\it flux-limited} sample expects $p\approx0$. Here, we find the posteriors to span values close to $0$, but with opposite sign to $x$, $p\approx-0.05$ to $-0.01$, which translates to finding slightly more sources the noisier the pixel rather than fewer. Causes for this might include Eddington bias or source confusion, although our current study is not conclusive on this. Nevertheless, the RMS correction chosen here effectively models the signal, which, especially for brighter flux density cuts, shifts the dipole directions' posteriors towards their expectation, and also restores the dipole amplitude's posteriors to be fully consistent with their respective expectations.

Finally, these results also illuminate the clustering present in the maps on different scales. The shaded bands around the kinematic dipole amplitude's expectation, $\mathcal{D}/\mathcal{D}_{\rm kin}=1$, indicate the possibility of a substantial clustering dipole in any one random sample given $\Lambda$CDM clustering (cf.~Table~\ref{tab:mocks1}). We therefore cannot expect the posteriors for our one mock to be centred on the exact expectation. Moreover, clustering matters for the inferred dipole parameters also on scales smaller than a dipole. While higher-order clustering multipoles are expected and further are known to correlate with the dipole parameters on a cut sky, it is the even smaller scales that lead to fluctuations on pixel scales, beyond just Poisson noise. In our likelihood, Eq.~\eqref{eq:lnl_genpoisson}, these are described effectively by the parameter $b$, which is inferred to lie around $b\approx0.5$ to $0.7$. This describes an increase in overall per-pixel variance by a factor of $1/(1-b)^2\approx4$ to $9$, correspondingly increasing the posterior widths. This is confirmed against a (RMS-corrected) Poisson likelihood, which overconfidently finds posterior widths consistent with the samples' shot noise expectations.

For an unbiased inference of the kinematic matter dipole present in the data, we find that redshift cuts are essential in distilling the kinematic from the clustering signal:  e.g.~at a flux density cut of $S_*=80\,\mu{\rm Jy}$, we find the expected clustering dipole's rms to be reduced from $\sigma_{\mathcal{D},{\rm cl}}/\mathcal{D}_{\rm kin}=55\%$ without redshift selection to $29\%$. For the latter choice we find $\mathcal{D}/\mathcal{D}_{\rm kin}=1.25^{+0.38}_{-0.33}$ -- a $\sim3.8\sigma$ detection of the dipole.  Further selections, if modelled properly into the kinematic expectation, may help reduce the clustering contribution further.  Higher precision for the dipole is only obtained if the small-scale clustering that leads to the $b$-inflated variance is modelled explicitly. Therefore, these results motivate the dedicated development of methods to incorporate all clustering signals into an appropriate inference pipeline, as well as more work regarding realistic simulations of all effects of the instruments and the cosmic large-scale structure to
identify the optimal survey designs for the various array configurations of SKA-Mid and SKA-Low in order to finally pin down the different components of the radio source count dipole.

\section{Kinematic effects on the power spectrum}
\label{sec:psdipole}

\subsection{The FOTO and B-FOTO signals}
\label{sec:psdipole_bfoto}

The Square Kilometre Array will probe the largest accessible cosmological scales through intensity mapping and HI galaxy surveys, allowing us to measure subtle relativistic and Doppler effects imprinted also on the clustering signal itself. Among these, the Finger of the Observer effect \citep[FOTO, the impact of the kinematic dipole on the 3D power spectrum multipoles, see e.g.][]{Elkhashab:2024yup,Bahr-Kalus:2021jvu,Bertacca:2019wyg} constitutes a distinct dipolar modulation of the three-dimensional power spectrum monopole caused by the observer velocity $\boldsymbol{\varv}_\mathrm{obs}$ with respect to the Cosmic Rest Frame (CRF) \citep[see also][]{Lacasa:2024ybp,Challinor:2011bk}\footnote{We avoid referring to this term as the power spectrum dipole to avoid possible confusion with the dipole of the Legendre expansion of the power spectrum, which, besides this term, contains other relativistic contributions \citep[see e.g.][]{Raccanelli:2013dza,Beutler_2020}.}. 
This signature reflects the same motion responsible for the kinematic CMB dipole and thus provides an independent, large-scale test of cosmic isotropy.
Building on the theoretical framework developed in \citet{Elkhashab:2024yup}, the FOTO effect appears as an oscillatory pattern in the monopole on very large scales ($k\lesssim 10^{-3}\,h/\mathrm{Mpc}$), with an amplitude proportional to $|\boldsymbol{\varv}_\mathrm{obs}|^2$.
The boosted FOTO signal (B-FOTO) introduced therein enhances its detectability by applying artificial redshift boosts in different directions, which modulate the monopole response and permit a joint reconstruction of the amplitude and direction of $\boldsymbol{\varv}_\mathrm{obs}$.
In this section we assess the ability of an SKA AA4 HI galaxy survey (as well as of a future extension thereof) to recover such a relativistic observer-motion imprint.

Following \cite{Elkhashab:2024yup}, it can be shown that the impact of the observer velocity $\boldsymbol{\varv}_{\rm obs}$ on the galaxy clustering statistics can be expressed as
\begin{equation}
\delta_\mathrm{obs}({\bf r})=\delta_\mathrm{CRF}({\bf r})+\alpha_0(z)\frac{\boldsymbol{\varv}_{\rm obs}\cdot \hat{{\bf r}}}{a(z) H(z) r}\,
\end{equation}
where $\delta_\mathrm{CRF}(\bf{r})$ and $\delta_\mathrm{obs}(\bf{r})$ are the galaxy over-densities computed at the comoving positions $\bf{r}$ by observers respectively comoving (i.e. at rest) and boosted by $\boldsymbol{\varv}_{\rm obs}$ with respect to the CRF. $H(z)$ and $a(z)$ are the Hubble parameter and scale factor evaluated at a redshift $z$, and $\alpha_0=\alpha_0(z)$ depends on the galaxy sample under study (not to be confused with the spectral index of Sec.~\ref{sec:numbercountsdipole}). 
For flux-limited tracers, such as the HI samples we consider in this analysis,
we can write the selection function as $\bar{n}_g(z) = \bar{n}\left(L_*(z),z\right)\,,$
where  $L_*(z)$ is the luminosity limit associated with a flux cut $S_*$ at a given redshift $z$.
Then, $\alpha_0$ reads
\begin{equation}
\label{eq:alpha_0}
    \alpha_0(z)=2(1-\mathcal{Q})+\left[3-\mathcal{E}-\frac{\mathrm{d}\,\mathrm{ln}\,H}{\mathrm{d}\,\mathrm{ln}\,(1+z)}\right]\frac{r\,H}{c(1+z)}\,,
\end{equation} 
where 
\begin{equation}
    \mathcal{E}(z) = - \frac{\partial\,\mathrm{ln}\,\bar{n}(L_*,z)}{\partial\,\mathrm{ln}\,(1+z)}\Bigg|_{L_*=L_*(z)}\, \quad 
{\rm and} \quad \quad
    \mathcal{Q}(z) = - \frac{\partial\,\mathrm{ln}\,\bar{n}(L_*,z)}{\partial\,\mathrm{ln}\,L_*}\Bigg|_{L_*=L_*(z)}
\end{equation}
are the evolution and magnification biases, as defined for a flux-limited sample, defined by the HI line flux.

Defining $P_{\mathrm{obs}}$ and $P_{\mathrm{CRF}}$ as the matter/galaxy power spectra of $\delta_\mathrm{obs}$ and $\delta_\mathrm{CRF}$, it can be shown that the impact of the FOTO effect on the power spectrum monopole in a full-sky survey is given by
\begin{equation}
\label{eq:foto}
    P_{0,\mathrm{dip}}(k) \equiv P_{0,\mathrm{obs}}-P_{0,\mathrm{CRF}} =\frac{16\pi^2}{3}\frac{\varv_\mathrm{obs}^2}{H_0^2}\frac{I_1^2(k, \alpha_0)}{4\pi\int_{r_1}^{r_2}r^2\bar{n}_\mathrm{g}^2\,\mathrm{d} r}\,,
\end{equation}
where 
\begin{equation}
\label{eq:I}
    I_1(k, \alpha) = \int_{r_1}^{r_2}\frac{r\,\bar{n}_\mathrm{g}\,\alpha}{a\,H/H_0}\,j_1(kr)\,\mathrm{d} r\,,
\end{equation}
with $j_\ell$ denoting the spherical Bessel function of the first kind of order $\ell$, and the integration limits correspond to the boundaries of the redshift bin considered. 

\citet{Elkhashab:2024yup} proposed a technique to increase the signal-to-noise of this signature and exploit it to constrain both the amplitude and direction in the sky of $\boldsymbol{\varv}_{\rm obs}$, by performing an artificial boost to the galaxies' heliocentric redshifts such that they transform as $ 1+z_{\rm art}(\hat{\bf r})= (1+z_{\rm hel})/(1+\boldsymbol{\varv}_{\rm art}\cdot \hat{{\bf r}}/c)$,
where $\boldsymbol{\varv}_{\rm art}$ is an arbitrary vector used to parametrize the boost.
Using the boosted redshifts to compute the galaxy radial distances they argued that more information can be extracted from the data set than from $P_{0,\mathrm{dip}}$ alone, as it can be shown that the impact of the B-FOTO signal on the power spectrum monopole is $P_{0, \rm \mathrm{B}dip}(k;\boldsymbol{\varv}_{\rm art}) \equiv  P_{0,\mathrm{B}} (k;\boldsymbol{\varv}_{\rm art}) - P_{0,\rm CRF} (k)$, where
\begin{equation}
\begin{split}
  P_{0, \rm \mathrm{B}dip}(k;\boldsymbol{\varv}_{\rm art})   &\approx \frac{16 \pi^2}{3H^2_0\int \bar{n}^2\,\mathrm{d}^3 r}
    \left\{\boldsymbol{\varv}_{\rm obs}^2 \,I_1^2(k,\alpha_0)\right.\\
    &\left.+2\,(\boldsymbol{\varv}_{\rm obs}\cdot \boldsymbol{\varv}_{\rm art})\,I_1(k,\alpha_0)[I_1(k,\alpha_0)+I_1(k,\Delta\alpha)]
    +\varv_\mathrm{art}^2 [I_1(k,\alpha_0)+I_1(k,\Delta\alpha)]^2\right\}\,,
 \end{split}
\label{eq:shifted_power_ang}
\end{equation}
and where
\begin{equation}
\label{eq:delta_alpha}
    \Delta \alpha=-\frac{2 r H}{c(1+z)}(\mathcal{Q}+1)\,.
\end{equation}

By performing several boosts, each time changing both the direction and amplitude of $\boldsymbol{\varv}_{\rm art}$, and measuring the resulting power spectrum monopole, one can attempt to constrain both the amplitude and direction of $\boldsymbol{\varv}_{\rm obs}$.
In particular, from Eq.~\eqref{eq:shifted_power_ang} we can construct a new observable, built by subtracting measurements of opposite boosts, such that
\begin{equation}
\label{eq:Op_boost}
\begin{split}
    \Delta P_{0,\mathrm{B}} & \equiv P_{0, \rm \mathrm{B}dip}(k;\boldsymbol{\varv}_{\rm art})-P_{0, \rm \mathrm{B}dip}(k;-\boldsymbol{\varv}_{\rm art}) \\
    &=\frac{64 \pi^2}{3H^2_0\int \bar{n}^2\,\mathrm{d}^3 r}
    \left\{
    (\boldsymbol{\varv}_{\rm obs}\cdot \boldsymbol{\varv}_{\rm art})\,I_1(k,\alpha_0)[I_1(k,\alpha_0)+I_1(k,\Delta\alpha)]
    \right\}\,,
\end{split}
\end{equation}
which retains information on both the amplitude and the direction of $\boldsymbol{\varv}_{\rm obs}$ through the product $\boldsymbol{\varv}_{\rm obs}\cdot \boldsymbol{\varv}_{\rm art}$.

\subsection{Mock catalogues}
\label{sec:psdipole_mocks}

We consider $3$ different $\mathrm{HI}$ galaxy samples, differing by the flux cuts $S_*=[5,120,150]\,\mu\mathrm{Jy}$.
The first cut describes an optimistic sample \citep[see e.g.][]{Maartens:2021dqy,SKA:2018ckk,Maartens:2015mra}, covering the redshift range $z\in[0.1,1.0]$, while the last two describe possible AA4 samples, covering the range $z\in[0.1,0.3]$.We will refer to these samples respectively as HI-5, HI-120 and HI-150.
The effects under study are relevant on ultra-large scales, where linear theory gives a sufficiently accurate description of the structure formation history. We thus rely on a linear biasing relation to construct our mock catalogues, via the \texttt{BUILDCONE} routine of \texttt{LIGER} \citep{Borzyszkowski:2017ayl,Elkhashab:2021lsk}. \texttt{BUILDCONE} exploits low-resolution DM-only Newtonian N-body simulations to generate galaxy number counts at the field level with the imprint of GR effects. It does so by ``painting'' the galaxy population on the DM density field obtained with \texttt{LIGER}, requiring in the process the selection function $\bar{n}_g(z)$, the linear bias $b_1(z)$ and the magnification and evolution biases $\mathcal{Q}(z)$, $\mathcal{E}(z)$ in the redshift bin under consideration.

To generate the required underlying matter density fields, we use third-order Lagrangian perturbation theory (3LPT), implemented by the \texttt{MONOFONIC} code \citep{2020ascl.soft08024H}.
For our simulation runs we consider a standard $\Lambda\mathrm{CDM}$ Planck 2013 cosmology \citep[][]{Planck:2013pxb} with $(\Omega_{\rm m}, \Omega_{\rm b}, \Omega_\Lambda, \sigma_8 , n_s , h) = (0.307, 0.048, 0.693, 0.829,0.96, 0.678)$.  Both this fiducial cosmology and the fiducial observer velocity adopted here differ slightly from those used in Sec.~\ref{sec:numbercountsdipole}, which are based on Planck 2018; these differences are immaterial at the level of precision of the present forecasts. 
We run two sets of simulations, each with the same number of particles ($256^3$), but with different box sizes, thus mass resolution. The first set is used for the AA4 tracers (HI-120 and HI-150), and presents a box size of $4000\,\mathrm{Mpc}/h$ which covers a full-sky light cone up to $z=0.8$ without box replications, and has a particle mass of $m_{\mathrm{dm}} = 3.25 \times 10^{14}\,\mathrm{M}_\odot/h$.
The second set is instead used for the extended sample (HI-5), and presents a box size of $8000\,\mathrm{Mpc}/h$, reaching $z=2.1$ in the full sky, and with a mass resolution $8$ times lower.
We then generate $60$ mock catalogues of each galaxy sample, accounting for all relativistic effects at linear order, including the FOTO effect, generated by a peculiar velocity of amplitude $\varv_\mathrm{obs} = \varv_\odot=369\,\mathrm{km}/\mathrm{s}$ and direction in the sky $(l_\odot,b_\odot)=(263.959^\circ, 48.260^\circ)$ as expected from the kinematic interpretation of the CMB dipole. 

For each run, we measure the boosted power spectrum signal of Eq.~\eqref{eq:Op_boost}, by performing opposite redshift boosts, with $\boldsymbol{\varv}_{\rm art}$ values taken from a set of amplitudes and directions in the sky listed in Table~\ref{tab:boosts}. We employ the Feldman--Kaiser--Peacock (FKP) estimator~\citep{Feldman:1993ky} to measure the power spectrum monopole
from the mock catalogues. Fourier transforms are computed after embedding the sections of the light cones within a cubic box  of side  $L_{\rm FFT} = 16 \,{\rm Gpc}/h$
and sampling the density on a Cartesian grid with $256^3$ voxels.
This setup allows us to measure $P_0$ for $k\geq k_\mathrm{f}=2\pi/L_{\rm FFT} = 3.9\times10^{-4}\,h/{\rm Mpc}$ and thus cover the range of wavenumbers for which the FOTO effect imprints a detectable signal. 
In order to compute the density contrast we estimate the mean local density within radial shells of thickness $\delta r = 10 \,\mathrm{Mpc}/h$. 
Since the scales of interest here lie far beyond those for which the FKP weights are designed, we simply set $w(r) = 1$.

\begin{SCfigure}
\centering
\includegraphics[width=0.5\textwidth]{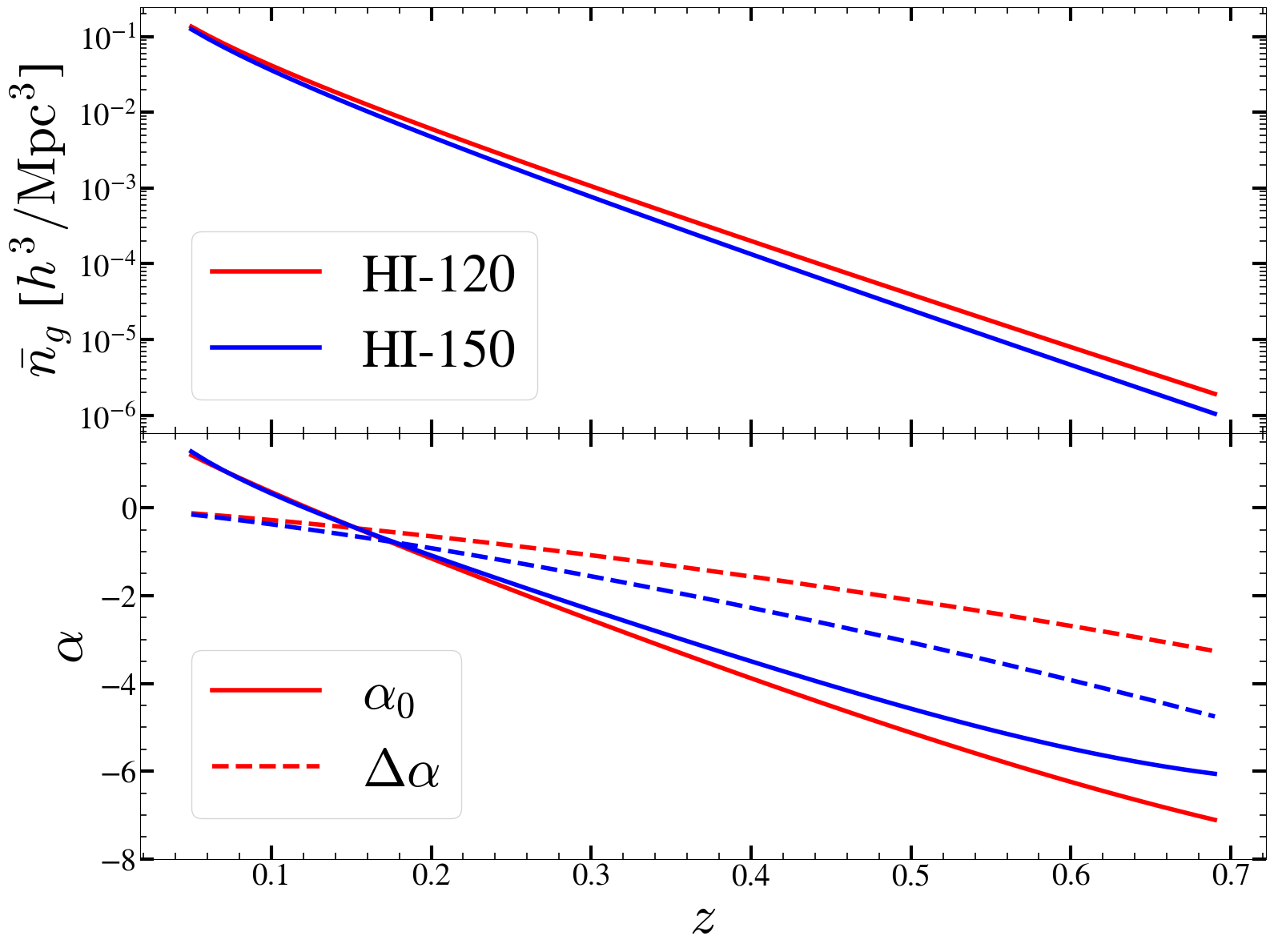}
\caption{\textit{Upper}: Selection functions of the two AA4-like tracers (HI-120 in red and HI-150 in blue). \textit{Lower}: Alpha functions (see Eqs.~\eqref{eq:alpha_0} and~\eqref{eq:delta_alpha}) for the same tracers. We show $\alpha_0$ (solid lines) alongside $\Delta\alpha$ (dashed lines). }
\label{fig:surv_funcs}
\end{SCfigure}

\begin{SCfigure}
\centering
\includegraphics[width=0.5\textwidth]{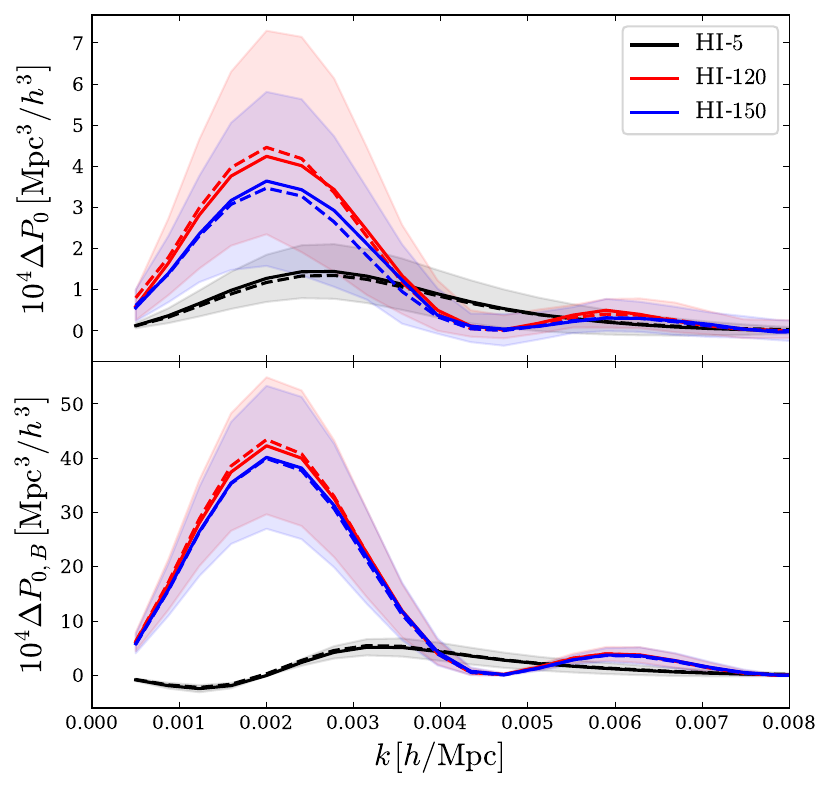}
\caption{ FOTO imprint on the power spectrum monopole for the HI-5 (black), HI-120 (red) and HI-150 (blue) samples. We show with shaded areas the $1\sigma$ scatter of the mock realizations (essentially cosmic variance) and with a solid line the average value. We plot for comparison the theoretical model of Eq.~\eqref{eq:foto} with dashed lines.  \textit{Upper:} Intrinsic FOTO contribution $\Delta P_0$. \textit{Lower:} Boosted signal $\Delta P_B$ of Eq.~\eqref{eq:Op_boost}.}
\label{fig:alpha}
\end{SCfigure}

In the upper panel of Fig.~\ref{fig:alpha}, we show the intrinsic FOTO contribution $\Delta P_0$ on the power spectrum monopole for the three samples under study. We consider a redshift bin $z \in [0.1,0.3]$ for the AA4 samples and a $z \in [0.1,1.0]$ bin for the extended HI-5 sample. In particular, regarding the AA4 samples, we avoided the redshift range in which $\alpha_0$ would change sign (see Fig.~\ref{fig:surv_funcs}), as it would decrease the value of the integral of Eq.~\eqref{eq:I}, while for the HI-5 case, we decided to encompass a larger volume to reduce cosmic variance in the constraints.
We plot the average value from the mocks (solid lines) and the $1\sigma$ scatter (shaded bands), compared with the theoretical expectation of Eq.~\eqref{eq:foto} (dashed lines). 
In the lower panel of the figure we instead show the same quantities for the boosted signal $\Delta P_\mathrm{B}$ of Eq.~\eqref{eq:Op_boost}, obtained by applying the third boost from Table~\ref{tab:boosts}. 

\begingroup
\renewcommand{\arraystretch}{1.2}
\begin{table}
\centering
\begin{tabular}{lr}
 \hline
 \textbf{Amplitude} $[\mathrm{km}/\mathrm{s}]$ & \textbf{Direction} $(l,b)\,[^\circ]$\\
 \hline
 $2 \varv_\odot$   & $(l_\odot,b_\odot)$\\
 $3 \varv_\odot$   & $(l_\odot+90^\circ,b_\odot)$\\
 $2 \varv_\odot$   & $(l_\odot,b_\odot-30^\circ)$\\
 $3 \varv_\odot$   & $(l_\odot+120^\circ,b_\odot)$\\
 \hline
\end{tabular}
\caption{Amplitudes and directions of the artificial boosts used in this analysis, expressed in terms of the fiducial values $\varv_\odot$, $l_\odot$, and $b_\odot$.}
\label{tab:boosts}
\end{table}
\endgroup

\sidecaptionvpos{figure}{t}

\sidecaptionvpos{figure}{t}

Finally, in this study, we estimate the survey functions of the galaxy populations following the approach of \citet{Yahya:2014yva}, based on the ${\rm S}^3$-SAX simulations~\citep{2009ApJ...703.1890O}. We measure the magnification bias $\mathcal{Q}$ from these simulations and infer the evolution bias $\mathcal{E}$ using the relations of \citet{Bertacca:2014hwa, Maartens:2021dqy}. While the ${\rm S}^3$ simulations underestimate the contribution of SFGs at low redshifts, as compared with, e.g.~the T-RECS simulations~\citep{Bonaldi:2023fnb} employed in Sec.~\ref{sec:numbercountsdipole}, an overall rescaling of $\bar{n}_g(z)$ within the low-redshift bin does not change $\alpha_0$ nor $\Delta\alpha$.
In the upper panel of Fig.~\ref{fig:surv_funcs} we show the selection functions of the AA4-like tracers (HI-120 in red and HI-150 in blue), while in the lower panel we show the $\alpha_0$ and $\Delta\alpha$ functions as solid and dashed lines, respectively.

\subsection{Parameter Inference}
\label{sec:psdipole_inference}

\begin{SCfigure}
\includegraphics[width=0.6\textwidth]{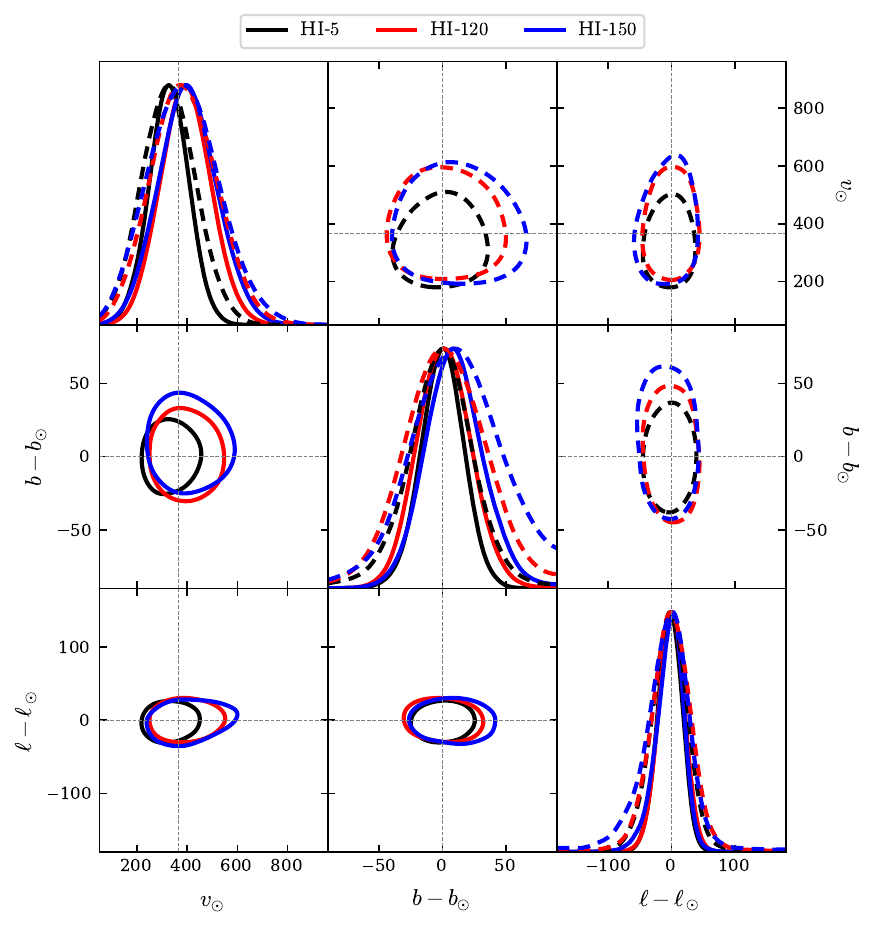}
\caption{Constraints on the amplitude ($\varv_{\rm obs}$) and direction ($l$, $b$) of our peculiar velocity obtained from the average of our mock realizations. We show respectively the constraints for a full-sky setting (\textit{bottom left}, full lines), and estimated corner plots for a half-sky setting (\textit{top right}, dashed lines). We show in black, red, and blue the constraints for respectively the HI-5, HI-120, and HI-150 samples.}
\label{fig:corner}
\end{SCfigure}

\begin{SCfigure}
\includegraphics[width=0.6\textwidth]{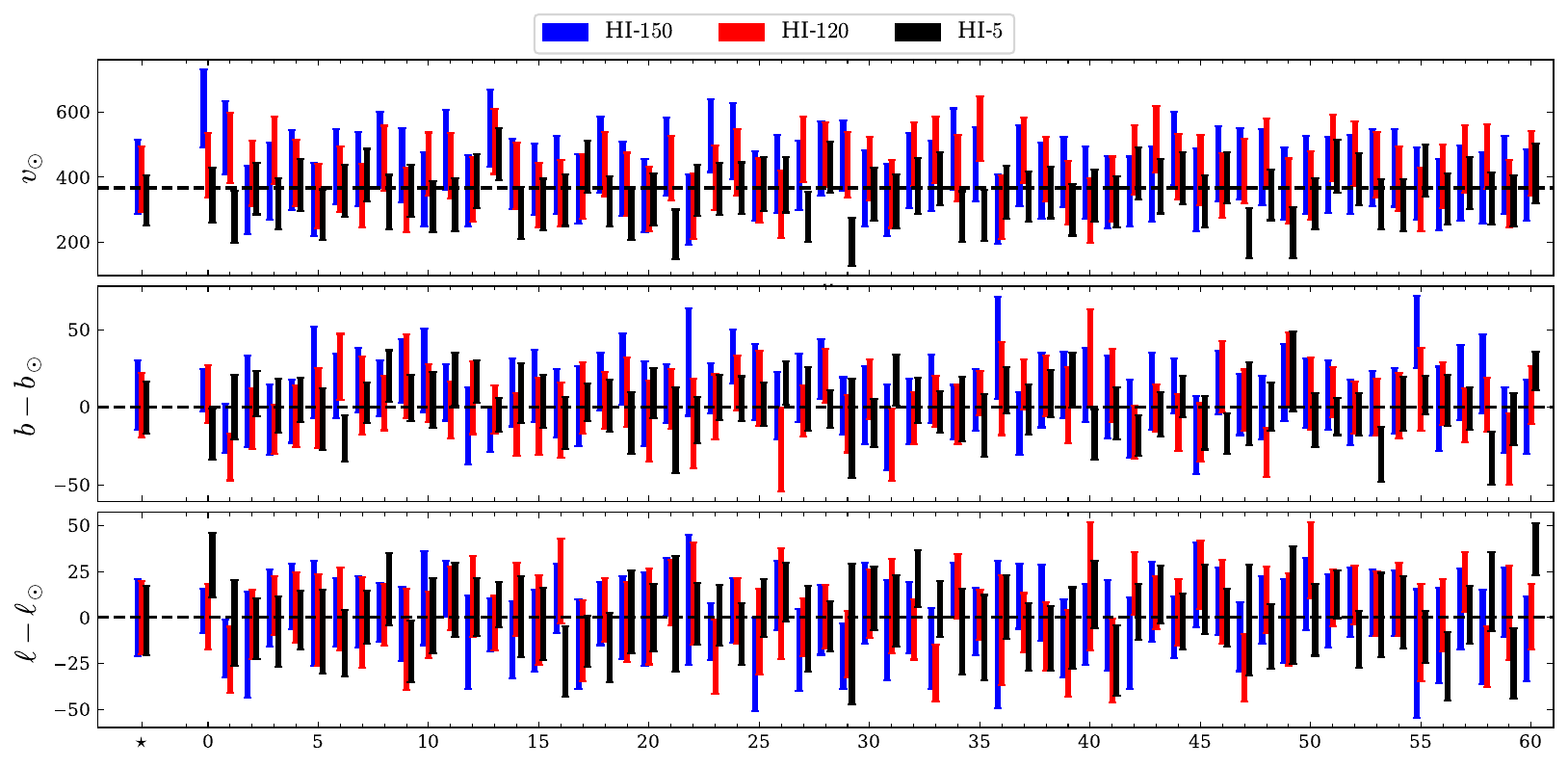}
\caption{Comparison of the $1\sigma$ constraints on $\varv_{\rm obs}$, $b$ and $l$ obtained from each one of the $60$ mock catalogues constructed for this work. We show in black, red, and blue the constraints for the HI-5, HI-120, and HI-150 samples  respectively. The error bars at the outermost left of the plot (associated with the $*$ symbol on the $x$-axis) show the constraints obtained using the average measurement of all $60$ realizations. With black-dashed lines we show the true values of the three parameters as set in this work.}
\label{fig:errorbars}
\end{SCfigure}

We implement the model of Eq.~\eqref{eq:Op_boost} and fit it to  measurements of the power spectrum monopole differences $\Delta P_{0,\rm{B}}$ as estimated from the mocks. We vary three parameters  $\vec{\theta}=[\varv_{\rm obs},b,l]$ representing the full vector of the velocity of the observer. 
For this analysis, we assume the survey functions are known and that the cosmology is set by the fiducial one. For the HI-120 and HI-150 surveys, we use the redshift bin $z \in [0.1,0.3]$ and use seven $k$ values, uniformly distributed in the range $8.2\times 10^{-4}\,h/\mathrm{Mpc}<k<20.8\times 10^{-3}\,h/\mathrm{Mpc}$. For the HI-5 measurements, we use the  $z \in [0.1,1.0]$ bin and seven $k$ values in the range $3.3\times 10^{-3}\,h/\mathrm{Mpc}<k<17.1\times 10^{-3}\,h/\mathrm{Mpc}$. We estimate the covariance from the scatter among the 60 mocks, after concatenating the measurements for the different boosts, and apply the Hartlap correction \citep{Hartlap:2006kj}.  Finally, we adopt wide uniform priors for the angular parameters over the whole sky and a Maxwell distribution for the $\varv_{\rm obs}$ with standard deviation $\sigma = 300\,{\rm km/s}$ (although the posteriors are not sensitive to this choice). We show the results for the average of the realizations in Fig.~\ref{fig:corner} on the bottom left triangle of the corner plot. In its upper right triangle, we show the results of the same analysis increasing the covariance by a factor of 2, corresponding to decreasing the volume of the respective survey by half. The $1\sigma$ interval constraints obtained for each parameter in the half-sky settings for the HI-150, HI-120, HI-5 samples are respectively $\sigma_{\varv_{\rm obs}}=[140.1,129.5,106.5]\,\mathrm{km}/\mathrm{s}$, $\sigma_{b}=[35.0,31.5,24.6]\,^{\circ}$ and $\sigma_{l}=[40.7,30.1,27.8]\,^{\circ}$.
We note that, while the FOTO and B-FOTO features of the HI-5 sample are the weakest among the samples considered, the constraints obtained for this extended sample are nevertheless the tightest, owing to the larger comoving volume probed.
For completeness, in Fig.~\ref{fig:errorbars}, we show the 1$\sigma$ constraints obtained in the full-sky case for each of the 60 mock catalogues created for each tracer, showcasing the measurements in the different realizations of the same cosmology. From this plot we confirm that each realization provides measurements which are consistent with the input values.
\sidecaptionvpos{figure}{t}

\sidecaptionvpos{figure}{t}

This analysis illustrates how large-scale measurements of the three-dimensional galaxy power spectrum can provide a new probe of our peculiar velocity, complementary to number-count dipole measurements. We show that through the FOTO signal it is possible to perform measurements of the observer’s motion by exploiting the population of HI galaxies at $z<0.5$, using samples compatible with those expected from AA4. Even tighter constraints could be achieved with a deeper HI galaxy survey extending to 
$z\sim 1-2$, as might be realized with a future expansion of the SKA.
Future efforts will need to focus on a detailed treatment of the survey geometry and on the mitigation of systematics that affect ultra-large scales, such as angular selection effects, which could otherwise bias the reconstruction of the FOTO signal and its cosmological interpretation.

\section{Kinematic effects on the spectra of radio backgrounds}
\label{sec:spectrumdipole}

In this section we consider yet another observable, upon which observer motion is imprinted, that will be measured by the SKA.  Just as for the CMB's spectrum, the spectra also of other radiation backgrounds (here dubbed `Cosmic Backgrounds', CB) appear modified to an observer that is moving with respect to them.  This leads to a dipolar imprint on the CB monopole's frequency spectrum, $\eta(\nu)$, assumed isotropic in its rest frame, and also leads to imprints of higher-order multipoles $\ell$ albeit with amplitudes decreasing as $\propto\beta^\ell$.  To first order in $\beta$, the CB dipole's spectrum (relative to the monopole's), expressed in terms of anisotropy of the equivalent thermodynamic temperature, is given by \citep{1981A&A....94L..33D}
\begin{align}
    \Delta T_{\rm th}(\nu) \simeq -\beta \, {h\,\nu \over k_{\rm B}\, (1+\eta(\nu))\, \ln^2(1+\eta^{-1}(\nu))}\,{{\rm d}\ln\eta(\nu)\over {\rm d}\ln \nu} \, ,
\end{align}
where $h$, $k_{\rm B}$ and $\eta(\nu)$ are the Planck and Boltzmann constants and the photon distribution function, respectively.
Various methods have been studied to compute these effects on CBs at different frequencies, and to detect the signal's frequency dependence and variation across the sky in different experimental setups~\citep[e.g.,][]{2011MNRAS.415.3227C,2014PhRvD..89l3504D,Balashev:2015lla,DeZotti:2015awh,Slosar:2016utd,Deshpande:2018ugh,Trombetti:2019oka,2024A&A...689A.294E}.  Accordingly, the CMB dipole's spectrum has been thoroughly investigated, e.g. as a check of the CMB monopole's spectrum or as a probe alternative or complementary to observations based on instrument absolute calibration~\citep[e.g.,][]{CORE:2017krr,Mukherjee:2018fxd}, in the context of both past and future experiments.
In this study, we forecast constraints on the amplitudes of the CB dipoles' spectra, as achievable with the SKA, which can provide another consistency test of the kinematic dipole signals discussed in the previous sections at different cosmic epochs.

\subsection{Sensitivity estimates within sky patches}

The cosmic radio background (CRB) as observable by SKA-Low and SKA-Mid consists of: the extragalactic source background (ESB), the cosmological free-free emission/distortion (FF), the redshifted 21 cm line (21-cm) from neutral hydrogen, the long wavelength tail of CMB spectral distortions, such as Bose-Einstein like distortions (BE), generated at high redshift, and Comptonization distortions (C), typically coupled to the FF, e.g.~via cosmic re-ionization.  Here, we focus on perspectives offered by the ESB which is the largest contribution to the CRB.  (For detailed discussion see~\cite{Trombetti:2025dfu} and~\cite{Burigana:2025bdm}.)

Since dipoles span the largest angular variation on the sky, it is commonly assumed that all-sky surveys or at least those with very large sky coverage are necessary for dipole analyses. However, signal variations in the sky due to a dipole pattern appear also on smaller scales, rendering questions about their detectability simply a matter of the signal-to-noise ratio.  As outlined in \cite{Trombetti:2019oka}, surveys with
a coverage 
of at least a few degrees especially along a meridian (in a reference frame with $z$ axis parallel to the dipole axis) 
are sensitive to a dipole, provided that they are carried out in regions sufficiently far from the dipole axis, near which the signal varies least, and from areas contaminated by Galactic foreground emission, e.g.~those at low Galactic latitudes.  A simple exploration based on the SKA's performance (as of~\cite{dewdney2016}) shows that this is a feasible approach~\citep[e.g.,][]{Trombetti:2025dfu,Burigana:2025bdm}.

The effect of the observer peculiar velocity on the frequency spectrum can be computed via the Compton-Getting effect~\citep[e.g.][]{1970P&SS...18...25F}, based on Lorentz invariance of the photon distribution function.  We adopt here theoretical predictions based on the formalism in \cite{Trombetti:2020fwp,Trombetti:2024jnq}, where explicit formulae up to $\ell = 6$ (i.e.~up to $\mathcal{O}(\beta^6)$) are given for arbitrary spectra, $\eta(\nu)$ as linear combinations of the signals in selected directions and the spectrum's derivatives.

\begin{SCfigure}
\includegraphics[width=0.68\textwidth]{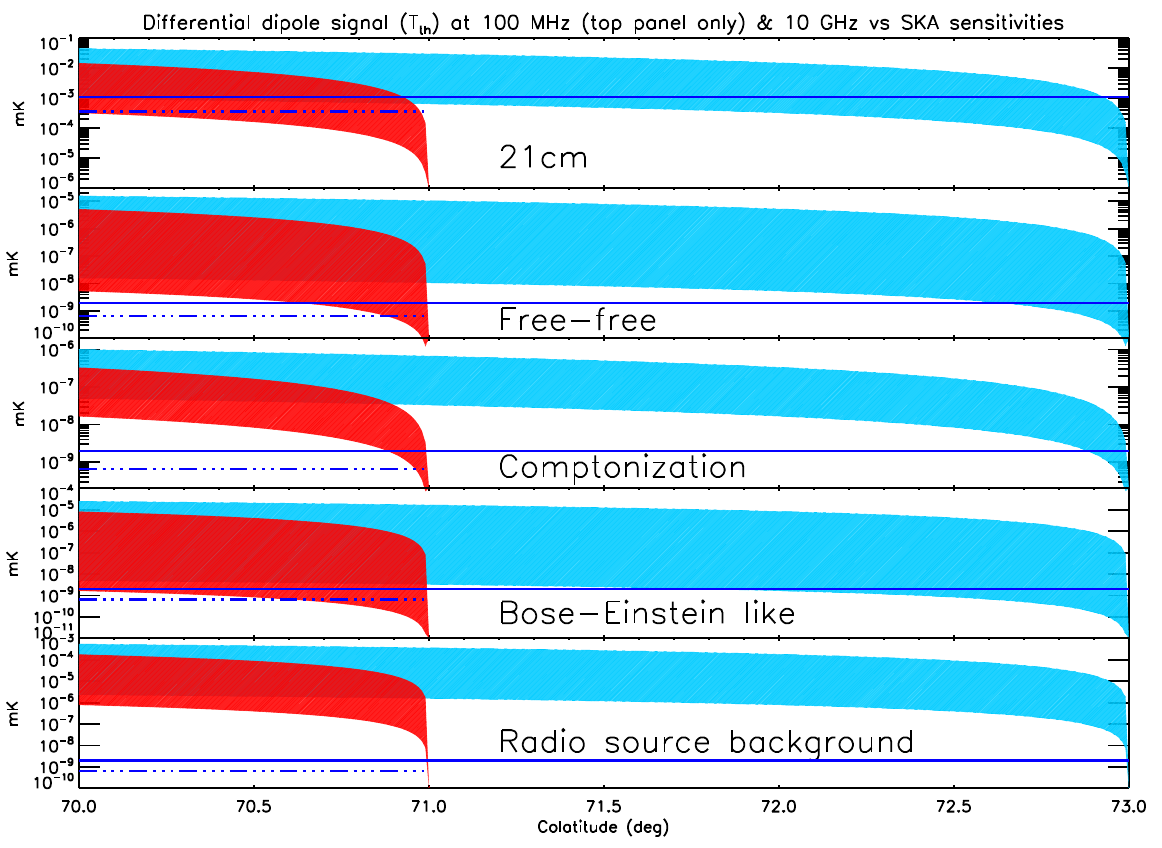}
\caption{
Differential temperatures of the most dominant CB contributions. 
We compare dipole temperatures $\Delta T(\theta) - \Delta T(\theta_{*})$ across patches of 3$^{\circ}$ (1$^{\circ}$) side length,  with $\theta_{*}$ = 73$^{\circ}$ (71$^{\circ}$).  The differential signals,
blue (red) bands, are compared with the SKA's sensitivity, solid (dash-dotted) lines.  Sensitivity estimates are based on 
\cite{dewdney2016} using a $2'$ side pixel and 1 day of integration time, and a 10\% (40\%) bandwidth for 21cm (other signals).
}
\label{fig:signals}
\end{SCfigure}

In Fig.~\ref{fig:signals} we show, in equivalent thermodynamic temperature, the ranges of expected contributions to the CB dipole spectrum, $\Delta T(\theta) = \Delta a_{1,0}\;[3/(4\pi)]^{1/2}\;{\rm cos} \theta$.  Here $\theta$ is the colatitude and $\Delta a_{1,0}$ refers to the dipole harmonic component $\ell=1$, $m=0$ in a frame oriented such that the $z$-axis is parallel to the observer motion.  Note that this is undertaken after the subtraction of the dipole spectrum corresponding to the fiducial CMB blackbody spectrum, in order to emphasize the CB's signals.  In each panel, the predicted differential signal's range is spanned by the top and bottom lines of the bands; for the ESB, the range also includes the subtraction of sources with fluxes above different detection thresholds (e.g.~$\gsim50\,{\rm nJy}$) --- see Fig. 5 in \cite{Trombetti:2025dfu} for a comparison at 100 MHz.  
The SKA's sensitivity turns out to be advantageous for the measurement of the CB's dipole contributions across a wide range of signal temperatures, even in small patches.

\sidecaptionvpos{figure}{t}

\subsection{Impact of Galactic foreground contamination}

To assess the SKA's capability to perform the described test, we aim to identify suitable sky regions, by characterizing their potential for accurate foreground removal.  To this end, we simulate the SKA's performance based on the SKAO sensitivity calculator as well as on assumed levels of (potential) residual Galactic foreground contamination.  Various methods of component separation, both blind and parametric \citep[see e.g.,][ and references therein]{2018JCAP...04..023R}, have been studied in this context, suggesting an accuracy in Galactic foreground removal at a level of a few percent.  Here, we rely on a parametric approach to quantify potential residuals from imperfect subtraction of Galactic foregrounds~\citep{CORE:2017krr}, and adopt the Global Sky Model~\citep[GSM,][]{Zheng:2016lul}, which provides $\sim\!{\rm arcmin}$ resolution based on a number of Galactic surveys across a wide range of frequencies.  We estimate sensitivity and resolution according to the AA4 configuration at $({\rm RA},{\rm Dec})=(51.6^{\circ}, -56^{\circ})$ using 20$^{\circ}$ (45$^{\circ}$) elevation for SKA-Low (SKA-Mid, assuming also a precipitable water vapour of 10 mm) and a Briggs' weighting of 0.  Sensitivity and resolution do not significantly degrade by adopting the AA* configuration.  The longer baselines are, however, crucial to mitigate source confusion noise, complementing the shorter baselines which are necessary for large-scale signal reconstruction.  Otherwise, without the longer baselines, the global sensitivity would be degraded by about two or three orders of magnitude, for the same pixel area.  We model the global signal in the pixel $p$ as
\begin{align}
    S(p) = T(p) + \big[ N_{\rm inst}(p) + N_{\rm conf}(p) \big] + \big[ E_{\rm fg}R(p) + E_{\rm fg,syst} \big] G(p) \, ,
\end{align}
including instrumental noise $N_{\rm inst}$, source confusion noise $N_{\rm conf}$ and two types of residuals from imperfect subtraction of the signal due to Galactic diffuse foreground $G(p)$.  Here, $(N_{\rm inst}(p) + N_{\rm conf}(p))$ is drawn from a random Gaussian with zero mean and a variance per pixel that accords with SKA specifications.  $E_{\rm fg}$ ($E_{\rm fg, syst}$) quantifies the amplitudes of the stochastic (systematic) residuals of imperfect Galactic subtraction, which we model to be proportional to the amplitude of the Galactic signal in $p$.  The stochastic residuals are modulated by $R(p)$, drawn from a random unit Gaussian.  Lastly, $T(p)$ stands for the CB signal that, in this case, includes the Doppler boosted radio ESB monopole \citep[cf.][]{2008ApJ...682..223G,Trombetti:2020fwp}, with corresponding kinematic terms up to $\ell = 6$.\footnote{Considering that the amplitudes of the kinematic terms scale roughly as $\beta^\ell$, Fig.~\ref{fig:signals} suggests that, under the ideal assumption of almost perfect foreground subtraction, the SKA sensitivity could be enough to detect at least the kinematic quadrupole.}

We consider circular patches of 100 deg$^{2}$, according to the size identified for the SKA-Low deep cosmological surveys in~\cite{SKA:2018ckk} (the sensitivity corresponding to the integration time used in Fig.~\ref{fig:signals} is conservative for SKA-Low and is comparable to the performance design of the SKA-Mid cosmology survey).  We distribute these evenly across the sky with pixel centres fixed to a HEALPix~\citep{Gorski:2004by} grid, with $N_{\rm{side}} = 64$ ($\simeq 0.92^\circ$ resolution).  The sky temperature is simulated with $N_{\rm{side}} = 512$ ($\simeq 6.9'$ resolution).  Within each patch we fit the signal $S(p)$ in terms of a monopole and dipole component and calculate the relative error on $\beta$ due to the presence of noise and residuals.  We
consider fits both at 100~MHz and at 1.355~GHz.

\begin{figure}
\centering
\subcaptionbox{100~MHz}{\includegraphics[width=0.48\textwidth]{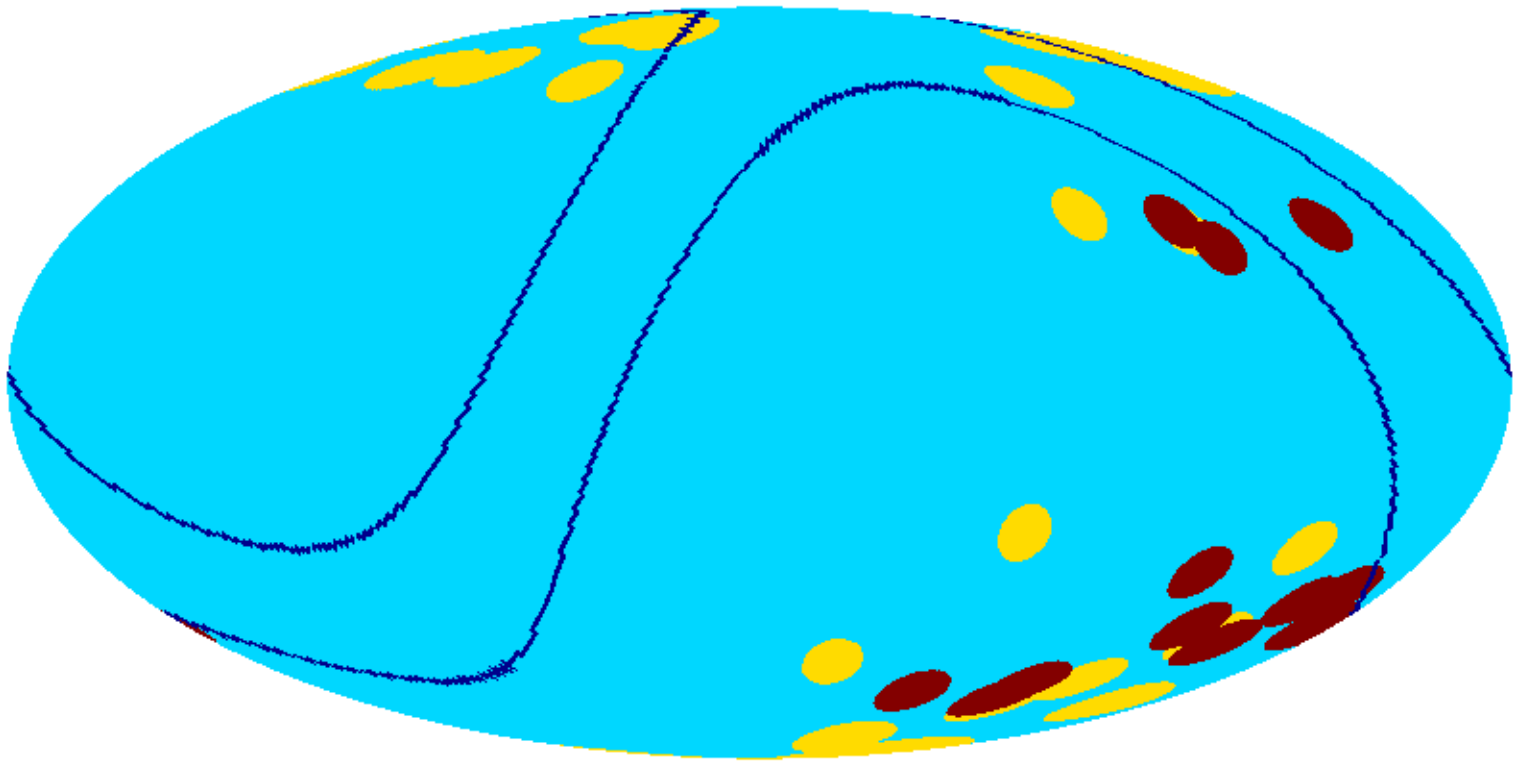}}
\subcaptionbox{1.355~GHz}{\includegraphics[width=0.48\textwidth]{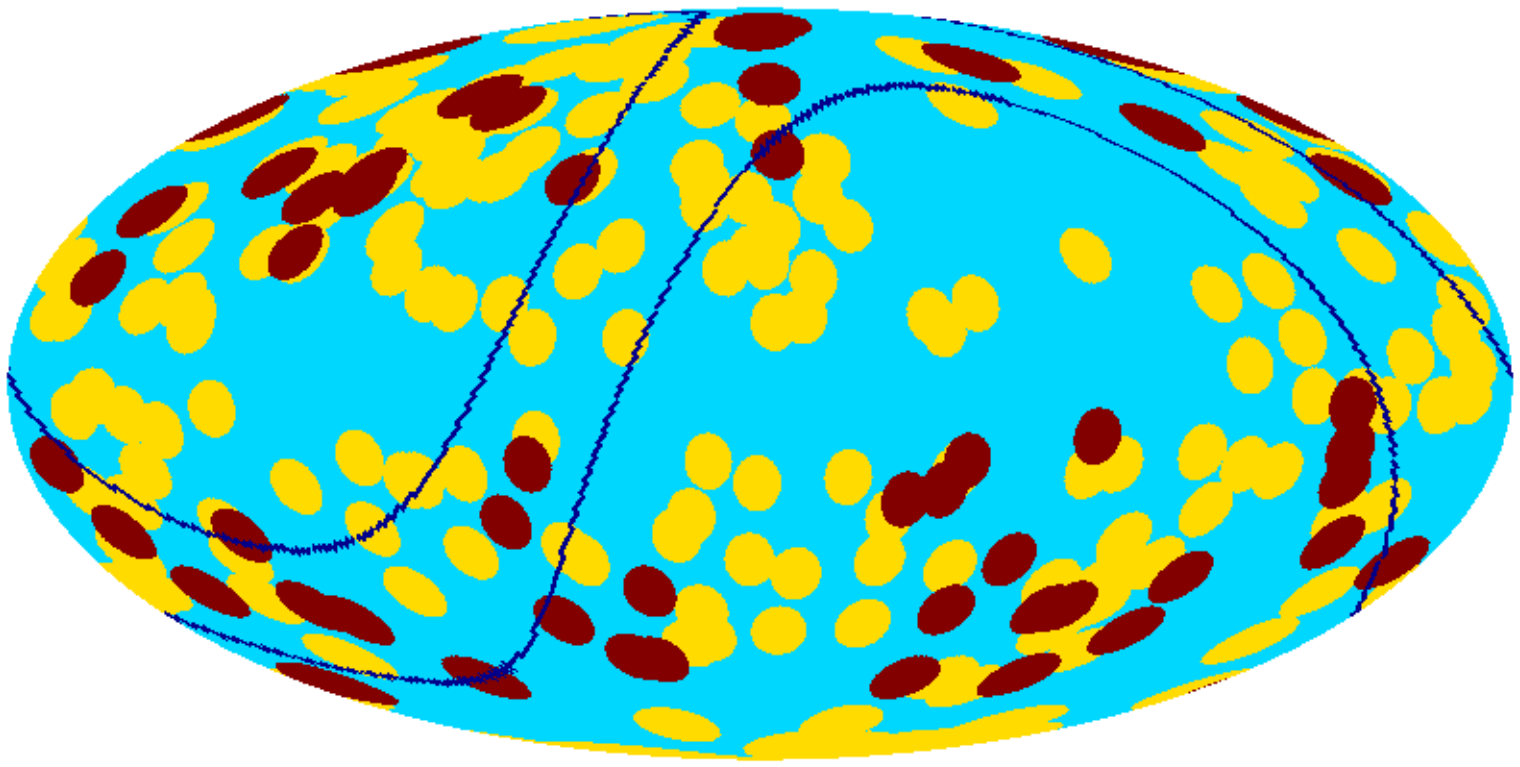}}
\caption{Selection of regions with relative uncertainty on the observer velocity $\beta$ of $<5\%$ (left) and $<50\%$ (right).  Regions are circular disks of size $100\,{\rm deg}^2$.  The foreground parameters used are $E_{\rm fg}=0.03$ and $E_{\rm fg,syst}=0.01$.  The blue lines indicate the range of possible maximum declinations for the SKA, ${\rm Dec}=30^\circ$ and ${\rm Dec}=0^\circ$.  In red the most favoured patches are shown, with relative uncertainties of $\lesssim3.5\%$ (left) and $\lesssim 25\%$ (right).}
\label{fig:patch}
\end{figure}

\begin{figure}
\centering
\subcaptionbox{100~MHz}{\includegraphics[width=0.48\textwidth]{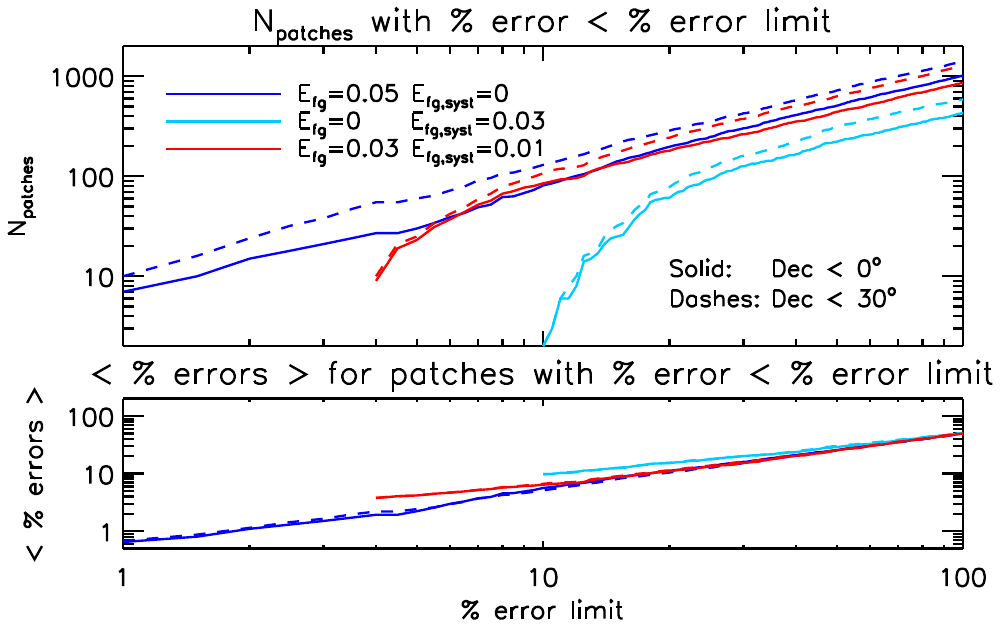}}
\subcaptionbox{1.355~GHz}{\includegraphics[width=0.48\textwidth]{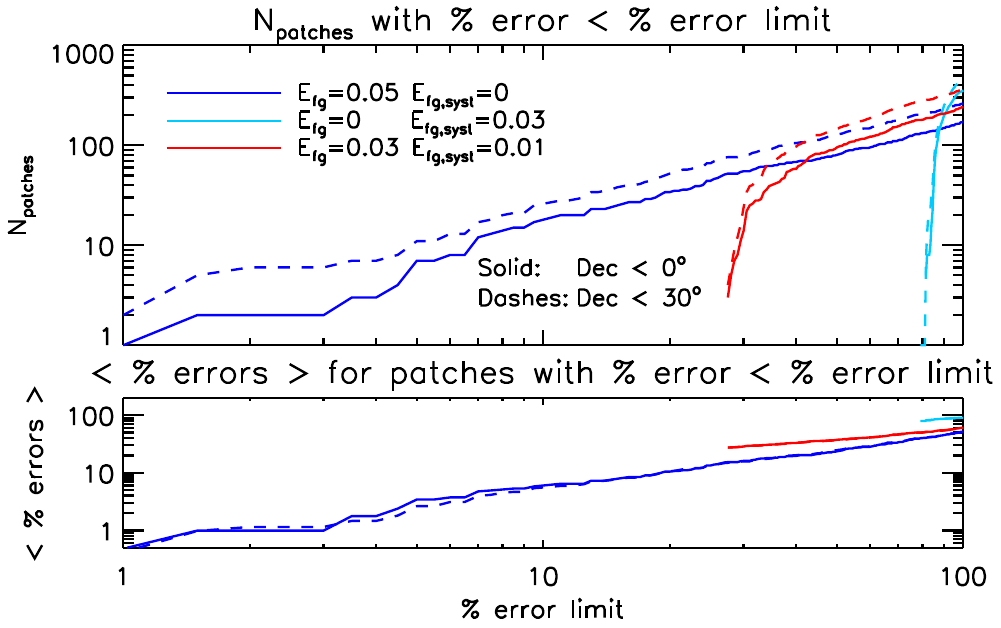}}
\caption{Top panels: cumulative distribution of numbers of patches with a relative error on $\beta$ below a certain limit. Dashed (solid) lines refer to patches at Dec$\lsim 30^\circ$ (Dec$\lsim 0^\circ$).  Three choices of values for $E_{\rm{fg}}$ and $E_{\rm{fg,syst}}$ are made (see legend).
Bottom panels: the corresponding mean across patches of the retrieved relative errors on $\beta$.}
\label{fig:patchnumbers}
\end{figure}

Considering only the stochastic Galactic residuals, $E_{\rm fg,syst}=0$, we find good agreement with the input value for many patches.  For example, for the conservative value $E_{\rm{fg}} = 0.05$ we find 60 (30) patches at ${\rm Dec}<30^\circ$ (${\rm Dec}<0^\circ$) at 100 MHz, all with a relative error on $\beta$ of $<5\%$ and, on average $\approx2.3\%$. At 1.355~GHz we find 26 (18) patches, with a relative error $<10\%$ and, on average $\approx5.8\%$.  In contrast, for $E_{\rm{fg,syst}}=0.05$, we find that coherent residuals correlated with the Galactic pattern represent the main obstacle and prevent satisfactory constraints on $\beta$.  Reducing the value to $E_{\rm{fg,syst}}=0.03$, we find 78 (61) patches at ${\rm Dec}<30^\circ$ (${\rm Dec}<0^\circ$) at 100 MHz, with a retrieved relative error on $\beta$ of $<20\%$ and, on average $\approx15\%$.  At 1.355~GHz we find 60 (40) patches, with a relative error $<85\%$ and, on average $\approx84\%$.  Thus, the requirement on a low $E_{\rm{fg,syst}}$ is more severe than that on $E_{\rm{fg}}$.  Finally, in Fig.~\ref{fig:patch}, we consider $E_{\rm{fg}} = 0.03$ along with $E_{\rm{fg,syst}} = 0.01$.  We show in the left (right) panel the best $100\,{\rm deg}^2$ patches within which measurements of $\beta$ to within $5\%$ (50\%) are predicted.  In addition, Fig.~\ref{fig:patchnumbers} shows the numbers of patches as a function of the limit of the retrieved relative error on $\beta$, for the three combinations of $E_{\rm{fg}}$ and $E_{\rm{fg,syst}}$ considered here, alongside the corresponding mean of the retrieved relative errors on $\beta$.

Tens of such patches (not significantly overlapping) are accessible to the SKA. Thus, even in the conservative case in which mosaic techniques and large scale cross calibration succeed only within scales comparable to the assumed patch scale, an average over the patches can reduce the error on $\beta$ to a few percent.  The ability to jointly analyse sky areas extending over a larger angular scale, allows us to work with a larger dipole differential signal and obtain a consequent improvement in relative accuracy --- of course under the provision that those patches are relatively free from Galactic foreground residual contamination.  This will be useful for radio ESB dipole investigations performed with decreasing source detection thresholds and, possibly, for exploring the subdominant components in the spectra of the Cosmic Backgrounds.

\section{Summary}
\label{sec:summary}

Measurement of our observer velocity with respect to cosmic backgrounds other than the microwave background offers a foundational check of the current standard model of cosmology as first discussed by~\citet{Ellis:1984uka}.  In this chapter we discussed three such measurements each of which is aimed at a different observable accessible to the SKA, and quantified the corresponding expected performance of the upcoming SKA surveys.

Studies of the~\citet{Ellis:1984uka} test with the SKA have long been anticipated for the large number of radio sources various surveys are expected to sample.  Initially, \citet{Schwarz:2015pqa} estimated percent-level uncertainties on recovered dipole amplitudes to be reached with SKA, based on simulations without gravitational clustering, but including flux calibration uncertainties. Subsequently, \citet{Bengaly:2018ykb} did include clustering and forecasted a $\sim5\sigma$ measurement of the expected kinematic matter dipole at a flux cut of $20\,\mu{\rm Jy}$, increasing to $\sim10\sigma$ when sources with $z<0.5$ were excluded, although no observational effects were considered there.  In contrast, our study includes both gravitational clustering and subtle observational effects, modelled in more detail than previously (see section 5 of ~\cite{Asorey01.2026.SKA} for a full description).  Via our simulations we found that, in addition to the kinematic signal, there are two previously unexplored contributions to the observed matter dipole that may have a large impact in upcoming data: the angular clustering by faint, low-redshift SFGs, and the anisotropic per-beam RMS.  The latter can be addressed by selecting sources with sufficiently high flux densities $S_{780\,{\rm MHz}}\gtrsim60\mu{\rm Jy}$ and by employing a suitable likelihood for parameter inference~\citep{Wagenveld:2023kvi}.  The former can be partially addressed via applying selections on observed redshifts.  These, in turn, produce an additional kinematic signal~\citep{vonHausegger:2024fcu}, which may be exploited in verifying the kinematic nature of the eventually observed matter dipole.  At the conservative flux density and redshift cuts of $S_{780\,{\rm MHz}}>80\mu{\rm Jy}$ and $z>0.5$, we expect to recover a dipole at a significance of $>3\sigma$, consistent with its kinematic expectation.  Upcoming data and methods will offer ample opportunity to sharpen such detections. Overall, an SKA-Mid Band 1 continuum survey would allow a very precise measurement of the combined kinematic and clustering source count dipole and provide what is necessary to put stringent upper limits on what fraction of that dipole might be due to the notorious kinematic observer boost.

As reported in Sec.~\ref{sec:psdipole}, SKA will enable measurements of subtle relativistic signatures in large-scale structure by probing ultra-large scales across the sky. Among these is the FOTO signal, a modulation of the galaxy power spectrum monopole arising from the observer's motion relative to the Cosmic Rest Frame. FOTO appears as an oscillatory feature on the largest scales, with an amplitude proportional to the square of the observer’s velocity, and its detectability can be enhanced through artificial redshift boosts (B-FOTO), which increase sensitivity to both the magnitude and direction of this velocity. Using mock HI galaxy catalogues of the AA4 setting and of a future, extended SKA HI survey—constructed with the LIGER framework, which generates relativistic light-cone simulations incorporating realistic selection, bias, and magnification functions—we show that boosted monopole measurements can recover the observer’s velocity vector from these samples across different sky fractions, with significantly tighter constraints achievable for the deeper, larger-volume extended (HI-5) sample. These results demonstrate that the FOTO signal provides a powerful and complementary probe of cosmic isotropy and observer motion, though fully leveraging this relativistic effect will require careful treatment of survey geometry and ultra-large-scale systematics.

Finally we explored in Sec.~\ref{sec:spectrumdipole} the possibility of measuring our observer velocity with respect to diffuse radio backgrounds.  We find that deep observations in $\sim100\,{\rm deg}^2$ patches, selected for their low Galactic foreground contamination, offer percent-level constraints on $\beta$ at 100~MHz, degrading to constraints at the level of tens of percent at 1.4~GHz. This estimate is robust for low values of systematic temperature uncertainty caused by imperfect Galactic foreground subtraction, which places demands on reliable foreground detection algorithms for SKA.

\section*{Author List Ordering}

Authors for this chapter are ordered alphabetically.

The corresponding authors are Sebastian von Hausegger and Dominik J.~Schwarz.

\section*{Acknowledgements}

DB acknowledges support from the COSMOS network (www.cosmosnet.it) through ASI (Italian Space Agency) Grants 2016-24-H.0, 2016-24-H.1-2018 and 2020-9-HH.0.  MYE is supported by the PRIN 2022 PNRR project ``Space-based cosmology with Euclid: the role of High-Performance Computing'' (code no. P202259YAF), funded by European Union – Next Generation EU.  CLH acknowledges support from the Oxford Hintze Centre for Astrophysical Surveys which is funded through generous support from the Hintze Family Charitable Foundation and from the Science and Technology Facilities Council (STFC) through grant ST/Y000951/1.  SvH is supported by a Leverhulme Trust International Professorship Grant to S. Sondhi (No. LIP-2020-014).  DJS acknowledges support by the German Federal Ministry for Research, Technology and Space under ErUM-Pro grant D-MeerKAT-III (05A23PBA).  MLS is supported by the Australian Government Research Training Program (RTP) Scholarship.  VM is supported by the University of Sydney’s Physics Foundation Scholarship.  OTO is supported by the University of Sydney Postgraduate Award.

Some of the results in this paper have been derived using the HEALPix \citep{Gorski:2004by} package. 

We also acknowledge the use of the computational resources of: Glamdring (University of Oxford); cuillin (Royal Observatory of Edinburgh); the University of Padova (through Strategic Research Infrastructure Grant 2017: ``CAPRI''); School of Physics HPC Cluster (School of Physics, University of Sydney).

\newpage
\bibliographystyle{AuxiliaryFiles/abbrvnat-maxbibnames4}
\bibliography{Bibliography}

\end{document}